\documentclass[10pt,aps,prl,twocolumn,superscriptaddress,preprintnumbers]{revtex4-1}
\pdfoutput=1
\usepackage{amsfonts}
\usepackage{mathrsfs}
\usepackage{amsmath}
\usepackage{amssymb}
\usepackage{fancyhdr}
\usepackage{graphicx}
\usepackage{xspace}
\usepackage{rotating}
\usepackage[normalem]{ulem}
\usepackage{braket}
\usepackage{verbatim}
\usepackage{xcolor}
\usepackage[utf8]{inputenc}
\usepackage[medium]{titlesec}
\usepackage{bm}
\usepackage[normalem]{ulem}
\usepackage{extarrows}
\usepackage{slashed}
\usepackage{isodateo}
\usepackage{graphicx}
\usepackage{xcolor}
\usepackage[bookmarksnumbered=true,bookmarksopen=true]{hyperref}
\usepackage[hmargin=.7in,vmargin=1.1in]{geometry}
\usepackage{indentfirst}
\usepackage{cancel}
\usepackage{soul}

\graphicspath{{./}{./Figs/}{./figs/}{../Figs/}{./fig/}}

\usepackage{amsfonts}
\usepackage{mathtools}
\usepackage{pifont}
\usepackage{mathrsfs}
\usepackage{amsmath}
\usepackage{amssymb}
\usepackage{framed}

\newcommand{\bit}{\begin{itemize}}  
\newcommand{\eit}{\end{itemize}}

\usepackage{accents}

\newcommand{\benum}{\begin{enumerate}}
\newcommand{\eenum}{\end{enumerate}}
\newcommand{\bi}{\begin{itemize}}
\newcommand{\ei}{\end{itemize}}
\newcommand{\Mp}{m_{\rm {Pl}}}

\newcommand{\beq}{\begin{equation}}
\newcommand{\eeq}{\end{equation}}

\newcommand{\bea}{\begin{eqnarray}}
\newcommand{\eea}{\end{eqnarray}}

\newcommand{\Rmnum}[1]{\expandafter\@slowromancap\romannumeral #1@}

\def\bga{\begin{aligned}}
\def\eda{\end{aligned}}
\def\bgp{\begin{pmatrix}}
\def\edp{\end{pmatrix}}
\def\bgs{\begin{subequations}}
\def\eds{\end{subequations}}

\def\to{\rightarrow}

\renewcommand{\rm}{\mathrm}

\newcommand{\es}[1]{{\textcolor{blue}{[es - {#1}]}}}

\usepackage{eso-pic}

\newcommand{\oi}[1]{\textcolor{purple}{[{\bf OI}: #1]}}

\begin{document}

\preprint{NORDITA-2025-030}

\title{Schwinger effect in axion inflation on a lattice}

\author{Oksana Iarygina}
\email{oksana.iarygina@su.se}
\affiliation{Nordita, KTH Royal Institute of Technology and Stockholm University, Hannes Alfv\'ens v\"ag 12, 10691 Stockholm, Sweden}
\affiliation{The Oskar Klein Centre, Stockholm University, 10691 Stockholm, Sweden}

\author{Evangelos I. Sfakianakis}
\email{evangelos.sfakianakis@austin.utexas.edu}
\affiliation{
Texas Center for Cosmology and Astroparticle Physics,
Weinberg Institute for Theoretical Physics, Department of Physics,
The University of Texas at Austin, Austin, TX 78712, USA}
\affiliation{Department of Physics, Harvard University, Cambridge, MA, 02131, USA}

\author{Axel Brandenburg}
\email{brandenb@nordita.org}
\affiliation{Nordita, KTH Royal Institute of Technology and Stockholm University, Hannes Alfv\'ens v\"ag 12, 10691 Stockholm, Sweden}
\affiliation{The Oskar Klein Centre, Stockholm University, 10691 Stockholm, Sweden}
\affiliation{McWilliams Center for Cosmology \& Department of Physics, Carnegie Mellon University, Pittsburgh, PA 15213, USA}
\affiliation{School of Natural Sciences and Medicine, Ilia State University, 3-5 Cholokashvili Avenue, 0194 Tbilisi, Georgia}

%

\begin{abstract}
We present the first lattice simulations of the nonlinear evolution after axion inflation by self-consistently incorporating currents arising from Schwinger pair production. The tachyonically amplified gauge fields trigger the growth of Schwinger currents, leading to universal values for the conductivity and magnetic field at the onset of strong backreaction and subsequent quenching of gauge field production.
We show that the Schwinger effect (prematurely) saturates gauge field production, thereby
diminishing the prospects of high scale axion inflation magnetogenesis as a viable solution for blazar observations.

\end{abstract}
\maketitle

\noindent\textbf{Introduction.} Inflation provides a compelling explanation for the origin of large-scale structure and the universe’s homogeneity and isotropy \cite{Guth:1980zm, Linde:1981mu, Albrecht:1982wi}. Among various models, axion (or natural) inflation \cite{Freese:1990rb, Svrcek:2006yi} is particularly appealing due to its theoretical robustness and natural embedding in UV-complete frameworks like string theory \cite{Baumann:2009ni, Blumenhagen:2014gta, Palti:2015xra, Czerny:2014xja, Higaki:2014pja, McAllister:2008hb}. The inflaton in these models is a pseudo-scalar axion-like field, whose shift symmetry ensures a flat potential across super-Planckian field excursions. The shift symmetry restricts the axion to derivative couplings, such as Chern-Simons terms ($\phi F\tilde F$), which lead to exponential gauge field amplification during and after inflation \cite{Anber:2009ua, Barnaby:2011qe}. These amplified modes can generate distinctive non-Gaussianities \cite{Barnaby:2010vf, Barnaby:2011qe, Barnaby:2011vw}, chiral gravitational waves \cite{Sorbo:2011rz, Cook:2013xea, Domcke:2016bkh, Garcia-Bellido:2016dkw, Bastero-Gil:2022fme, Garcia-Bellido:2023ser, Corba:2024tfz}, primordial magnetic fields \cite{Garretson:1992vt, Anber:2006xt, Fujita:2015iga, Adshead:2016iae, Durrer:2023rhc}, and nearly instantaneous preheating \cite{Adshead:2015pva, Cuissa:2018oiw, Adshead:2023mvt}. These phenomena have been revisited in recent lattice simulations \cite{Adshead:2019igv, Adshead:2019lbr, Adshead:2023mvt, Caravano:2021bfn, Caravano:2022epk, Figueroa:2023oxc, Caravano:2024xsb, Sharma:2024nfu, Figueroa:2024rkr}.

Moreover, the induced electric fields can become strong enough to trigger Schwinger pair production \cite{Sauter:1931zz, Heisenberg:1936nmg, Schwinger:1951nm, Kluger:1991ib}. Due to its nonlinear and nonperturbative nature, accurately capturing this effect in axion inflation has drawn considerable attention, motivating various proposals \cite{Domcke:2018eki, Sobol:2019xls, Domcke:2019qmm, Gorbar:2021rlt, Fujita:2022fwc, Gorbar:2023zla, vonEckardstein:2024tix}.

This Letter contains results from the first lattice simulation of preheating after axion inflation that self-consistently includes the Schwinger effect. We demonstrate a suppression of the produced electric and magnetic fields,  effectively ruling out primordial magnetogenesis from high scale axion inflation\footnote{In this context,  high-scale inflation refers to  models where the inflaton potential can be well approximated by a quadratic near the end of inflation with $m^2_\phi=V''(\phi) \simeq (10^{-6}m_{\rm Pl})^2$ and furthermore a Hubble scale at the end of inflation which is similar to $m_\phi$.}. We discover a universal value for both the electromagnetic (EM) fields as well as the conductivity of the Schwinger plasma at the onset of backreaction and present a simple derivation of these values, based on the competition between the axion-gauge coupling term and the Schwinger current in the equation of motion of the electric field. 
Furthermore, we explore the effect of the mass of the lightest Standard Model fermions and show that a large Higgs vacuum expectation value (VEV) during inflation can weaken the Schwinger backreaction, partially restoring gauge field growth.

The rest of the Letter is organized as follows. We start by presenting the basics of the model and the different descriptions of the Schwinger-induced current. Following that, we present the results of our numerical simulations and analytic estimates. We conclude with the limitations of our method and outlook for future work.\\

\noindent\textbf{Axion inflation and the Schwinger effect.}
We consider a pseudoscalar inflaton (axion) $\phi$ 
coupled to the hypercharge sector of the Standard Model through a Chern-Simons interaction term in the presence of charged particles 

\begin{eqnarray}
\nonumber
        S=\int d^4 x \sqrt{-g} \left [-\frac{1}{2} \partial_{\mu}\phi\partial^{\mu}\phi - V(\phi) \right .
        \\
       \left .
        -\frac{1}{4}F_{\mu\nu}F^{\mu\nu}-
	\frac{\alpha}{4f}\phi F_{\mu\nu}\tilde{F}^{\mu\nu} +{\cal L}_{\rm {ch}} \right ] \, ,
    \end{eqnarray}
where $F_{\mu\nu}=\partial_{\mu}{A}_{\nu}^{\rm {ph}}-\partial_{\nu}{A}_{\mu}^{\rm {ph}}$, $\alpha$ is the axion-gauge coupling, $f$ is the axion decay constant, $V(\phi)$ is an axion potential and ${\cal L}_{\rm {ch}}={\cal L}_{\rm {ch}}({A}_{\mu}^{\rm {ph}}, \chi)$ describes all charged fields, $\chi$, and their interaction with $A_{\mu}^{\rm {ph}}$. With the superscript ``${\rm {ph}}$" we denote \textit{physical} fields.
The physical electric four-current is then
$ J^{\mu}=-{\partial {\cal L}_{\rm {ch}}}/{\partial A_{\mu}}=\left(\rho_{\rm{ch}}, {\bm{J}}^{\rm {ph}}/a\right)$, where $a(t)$ is the scale factor and we assume charged particles initially absent (or exponentially diluted during inflation)  and thus set the initial charge density to zero, $\rho_{\rm{ch}}=0$. 
It is convenient to work with 
 \textit{comoving} fields that relate to physical as $\bm{E}=a^2 {\bm E}^{\rm {ph}}$, $\bm{B}=a^2 {\bm B}^{\rm {ph}}$, $\bm{J}=a^3 {\bm{J}}^{\rm {ph}}$. 
Comoving  electric and magnetic fields are defined as
$\bm{E}=-\partial_\tau \bm{A}+\bm{\nabla}A_0$, $\bm{B}=\bm{\nabla}\times \bm{A}$, where we use derivatives with respect to conformal time $d\tau=dt/a(t)$.
The dynamical equations that govern the evolution of the (comoving) gauge and axion fields are \cite{Gorbar:2021rlt, Domcke:2018eki, Domcke:2019qmm}
\begin{align}
&\partial_\tau^2\phi+2{\cal{H}}\partial_\tau\phi-\bm{\nabla}^2\phi +a^2\frac{dV}{d \phi} =\frac{\alpha}{ a^2 f}\bm{E}\cdot\bm{B}, \label{eq:phi}\\
& \partial_\tau\bm{E}-{\rm{rot}}\, \bm{B}+\frac{\alpha}{f}\left(\partial_\tau\phi\bm{B}+\bm{\nabla}\phi\times \bm{E}\right)+\bm{J}=0,\label{eq:Edotconf}\\
&\bm{\nabla}\cdot \bm{E}=-\frac{\alpha}{ f}\bm{\nabla}\phi\cdot \bm{B}, \quad \bm{\nabla}\cdot \bm{B}=0,\\
&\partial_\tau{\bm{B}}+{\rm{rot}}\,\bm{E}=0,\\
&{\cal H}^2=\frac{8\pi}{3m_{\rm {Pl}}^2}a^2\left(\rho_{\phi}+\rho_{E}+\rho_{B}+\rho_{\chi}\right), \label{eq:H}
\end{align}
where ${\cal H}=\partial_\tau a/a$ is the conformal Hubble parameter. The physical energy densities are 
$\rho_{\phi}=\left \langle (\partial_\tau \phi)^2/2 a^2+(\bm{\nabla}\phi)^2/2 a^2+V \right \rangle$  for the axion, $\rho_E=\left \langle \bm{E}^2\right \rangle /2a^4$ for the electric field,   $\rho_B=\left \langle \bm{B}^2\right \rangle /2a^4$ for the magnetic field, and $\rho_{\chi}$ for the plasma. In the simulation, $\langle ...\rangle$ denotes volume averaging over the whole simulation domain. We use the temporal gauge, $A_0=0$.\\

\noindent\textbf{Strong backreaction  from Schwinger currents.}
The induced Schwinger current generated by the created particles for the case of constant and spatially uniform (anti)collinear electric and magnetic fields in de Sitter space
takes the form \cite{Bavarsad:2017oyv, Domcke:2018eki, Domcke:2019qmm, vonEckardstein:2024tix}
\begin{equation}
  J=  \frac{(e |Q|)^3}{6\pi^2 {\cal H}}E|B| \coth\left( \frac{\pi |B|}{E}\right)e^{-\frac{\pi m^2a^2}{e |Q|E}}, \label{eq:current}
\end{equation}
where $E=|\bm{E}|$ is the magnitude of the electric field and $J$, $B$ are the electric current and magnetic field, projected onto the direction of the electric field, $e$ is the gauge coupling constant, $Q$ is the particle’s charge, and $m$ is the particle's mass. We  focus our attention on the strong-field limit, defined as \cite{vonEckardstein:2024tix} $|e Q E|\gg {\cal H}^2$, meaning that we choose couplings that would generate an $E$-field that satisfies the above inequality in the absence of a Schwinger plasma. 
We also neglect the fermion masses by assuming $m \pi a^2 \ll e |Q|E$, unless otherwise stated. When electric and magnetic fields are (anti-)collinear, the induced current is proportional to both $\bm{ E}$ and $\bm{B}$.
This results in an ambiguity in writing a vector form for the Ohm’s law for the Schwinger current and 
allows for different formulations,  dubbed the ``electric", ``magnetic" and ``mixed" pictures,
\begin{align}
& \bm{J} = \sigma_E \bm{E},  \quad
\sigma_E = \frac{(e |Q|)^3}{6\pi^2 \mathcal{H}} |B| \coth\left( \frac{\pi B}{E} \right), \label{eq:sigmaE} \\
& \bm{J} = \sigma_B \bm{B}, \quad
\sigma_B = \frac{(e |Q|)^3}{6\pi^2 \mathcal{H}} \text{sign}(\bm{E}\cdot\bm{B}) \, E \coth\left( \frac{\pi B}{E} \right), \label{eq:sigmaB} \\
 & \bm{J} = \sigma_E \bm{E} + \sigma_B \bm{B}, \label{eq:mixed}
\end{align}
respectively,
where for the mixed picture the conductivities $\sigma_{E}$, $\sigma_{B}$ are chosen to satisfy Eq.~\eqref{eq:current}. We refer to this description of conductivities as \textit{collinear}, to emphasize the underlying assumption of (anti-)collinearity of the fields.

However, in axion inflation, electric and magnetic fields may not remain collinear or anti-collinear at all times. Relaxing the assumption of collinearity was addressed by performing a Lorentz boost from the comoving coordinate frame to a frame in which the electric and magnetic fields are collinear, and then transforming back. This was first explored perturbatively by considering small deviations around constant, anti-collinear background fields in Ref.~\cite{Fujita:2022fwc}, and later extended to a non-perturbative treatment in Ref.~\cite{vonEckardstein:2024tix}. Since no assumption is made about the collinearity of the fields, and they can take arbitrary configurations, we refer to this description as \textit{non-collinear}. This procedure leads to the induced current in the mixed picture \eqref{eq:mixed}, where it is described  through both an electric and magnetic conductivities as \cite{vonEckardstein:2024tix}
\begin{equation} \label{eq:noncolsigma}
        \sigma_E = \frac{|\bm{J}'|E'}{\gamma \, I^2}\,  , \quad 
    \sigma_B = \frac{|\bm{J}'|}{E' \gamma\, I^2}(\bm{E}\cdot \bm{B}),
\end{equation}
where $I^2 \equiv \sqrt{(\bm{E}^2-\bm{B}^2)^2+4(\bm{E}\cdot \bm{B})^2}$ and
 prime quantities are fields in the collinear frame 
defined through an arbitrary configuration of comoving $\bm{E}$ and $\bm{B}$ fields as
\begin{align}
J'&=  \frac{(e |Q|)^3}{6\pi^2 { \cal{H}}}E' |B'|\coth\left( \frac{\pi |B'|}{E'}\right), \\
    E'&=\frac{1}{\sqrt{2}}\left[\bm{E}^2 -\bm{B}^2+I^2\right]^{1/2},\\
    B'&=\frac{{\rm{sign}}(\bm{E}\cdot \bm{B})}{\sqrt{2}}
    \left[\bm{B}^2 -\bm{E}^2+I^2\right]^{1/2}, \\
    \gamma&=\frac{1}{\sqrt{2}}\left[1+ \frac{\bm{E}^2+\bm{B}^2}{I^2}\right]^{1/2}\, .
\end{align}
To obtain a closed system of equations, one needs to account for the evolution of the physical fermion energy density, $\rho_{\chi}$. Incorporating energy conservation in an expanding universe, the equation for $\rho_{\chi}$ can be written phenomenologically as \cite{Sobol:2019xls,vonEckardstein:2024tix}
\begin{equation}\label{eq:fermion_energy}
    \partial_\tau\rho_{\chi}+4 {\cal H} \rho_{\chi}=\frac{1}{a^4}\left(\langle\sigma_E \rangle\, \langle \bm E^2\rangle+\langle\sigma_B \rangle \langle \bm{E}\cdot \bm{B}\rangle\right),
\end{equation}
where it is assumed that the plasma is comprised of relativistic particles possessing a statistically isotropic momentum distribution, $p_{\chi}=\rho_{\chi}/3$.  It is worth noting that in Eq.~\eqref{eq:fermion_energy} 
one could use $\langle\sigma_E {\bm E}^2\rangle$+ $\langle\sigma_B  \bm{E}\cdot \bm{B}\rangle$.
We defer a detailed comparison of different prescriptions for a subsequent publication.
\\

\noindent\textbf{Numerical simulations.}
To determine the evolution of the system, we solve Eqs.~\eqref{eq:phi}--\eqref{eq:H} together with Eq.~\eqref{eq:fermion_energy}. This is done numerically on a lattice using the {\sc Pencil Code}~\cite{PencilCode:2020eyn}.
For our choice $\alpha m_{\rm {Pl}}/f=60$, a grid of $512^3$ points is sufficient. We start around $3$ $e$-folds before the end of inflation and initialize fields with the Bunch-Davies initial conditions.
For simplicity we choose a quadratic potential for the axion $V(\phi)=\frac{1}{2}m^2\phi^2$ with $m=1.04 \times 10^{-6}\, m_{\rm {Pl}}$.
Even though this is observationally ruled out during inflation, it is a valid approximation during preheating and as such has been widely used in the literature \cite{Figueroa:2023oxc, Caravano:2024xsb,  Figueroa:2024rkr, Sharma:2024nfu}. We do not expect qualitative differences for more complicated potentials, like axion monodromy~\cite{Adshead:2015pva}.

As outlined above, there are several possible descriptions of the Schwinger current. 
This is  due to the non-perturbative nature of the effect and the fact that the solution is only known for a constant electric field limit.
Thus extrapolating from this to more realistic scenarios leads to different prescriptions.
We begin our analysis with the simplest parametrization: the collinear current description in the electric picture, given by Eq.~\eqref{eq:sigmaE},  and compare it with the non-collinear formulation Eq.~\eqref{eq:mixed} -- \eqref{eq:noncolsigma}.
We follow the definitions of Ref.~\cite{vonEckardstein:2024tix} for the charge and the gauge coupling constant. In the expression for the conductivities we set  $Q^3=41/12$, which equals half the sum of the cubes of the hypercharges of all Standard Model Weyl fermions (while Eq.~\eqref{eq:current} refers to a single Dirac fermion). 
The gauge coupling constant is $e=g'=\sqrt{4\pi/137}\simeq 0.303$, but a realistic description of the Schwinger effect requires taking into account its running.
Hence in our simulations we use the gauge coupling constant $e= g'(\tilde{\mu})$ defined as
\begin{equation}
    g'(\tilde{\mu})=\left([g'(m_Z)]^{-2}+\frac{41}{48 \pi^2}\ln\frac{m_Z}{\tilde{\mu}} \right)^{-1/2},
\end{equation}
where $g'(m_Z)\simeq 0.35$, $m_Z\simeq 91.2 \, \rm{GeV}$, with the characteristic energy scale $ \tilde{\mu}=(\rho_E+\rho_B)^{1/4}$.
The conductivities in Eqs.~\eqref{eq:sigmaE} and \eqref{eq:noncolsigma} depend  on the electric and magnetic fields.
In our numerical simulations, the conductivity is computed \textit{locally} from the fields at each grid point. Nevertheless, defining it in terms of spatially averaged fields yields nearly identical results (detailed comparison will be presented in a subsequent publication).

\begin{figure}[h!]\begin{center}
\includegraphics[width=\columnwidth]{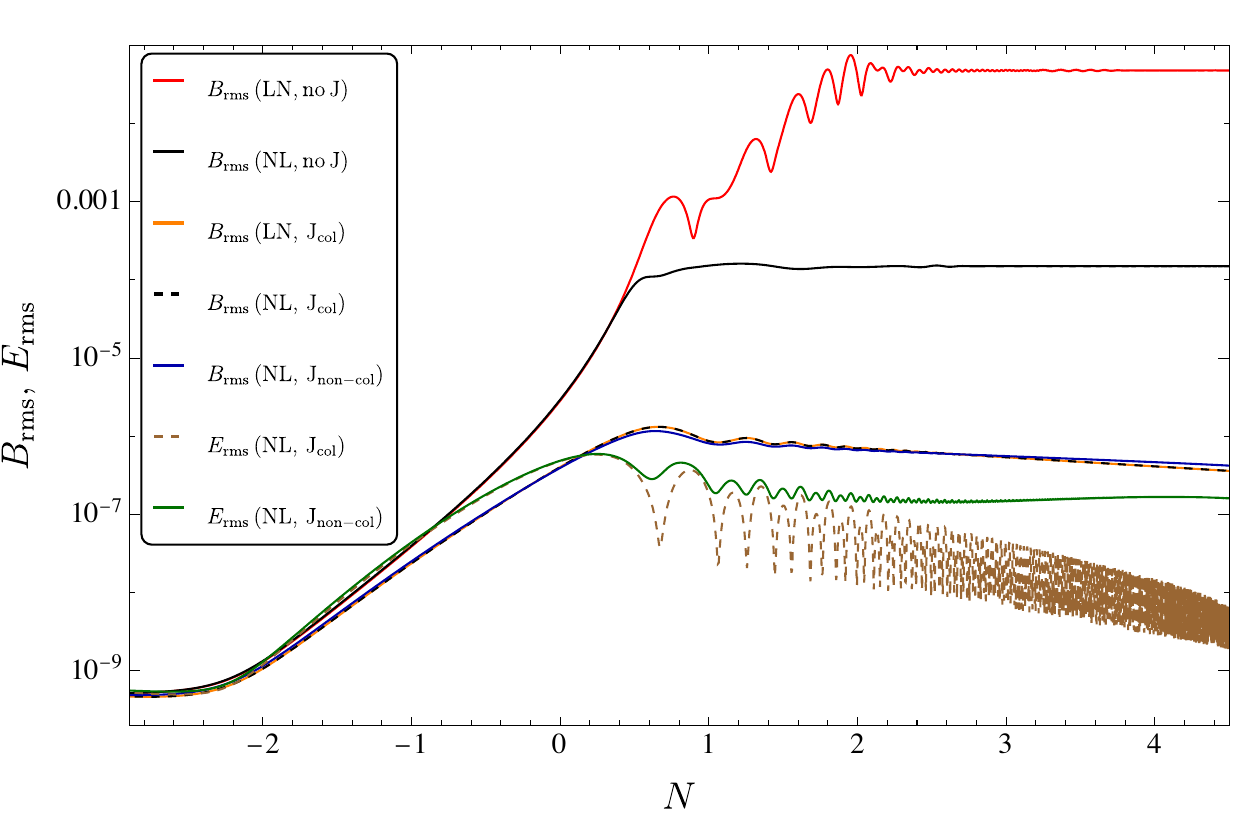}\\
\includegraphics[width=\columnwidth]{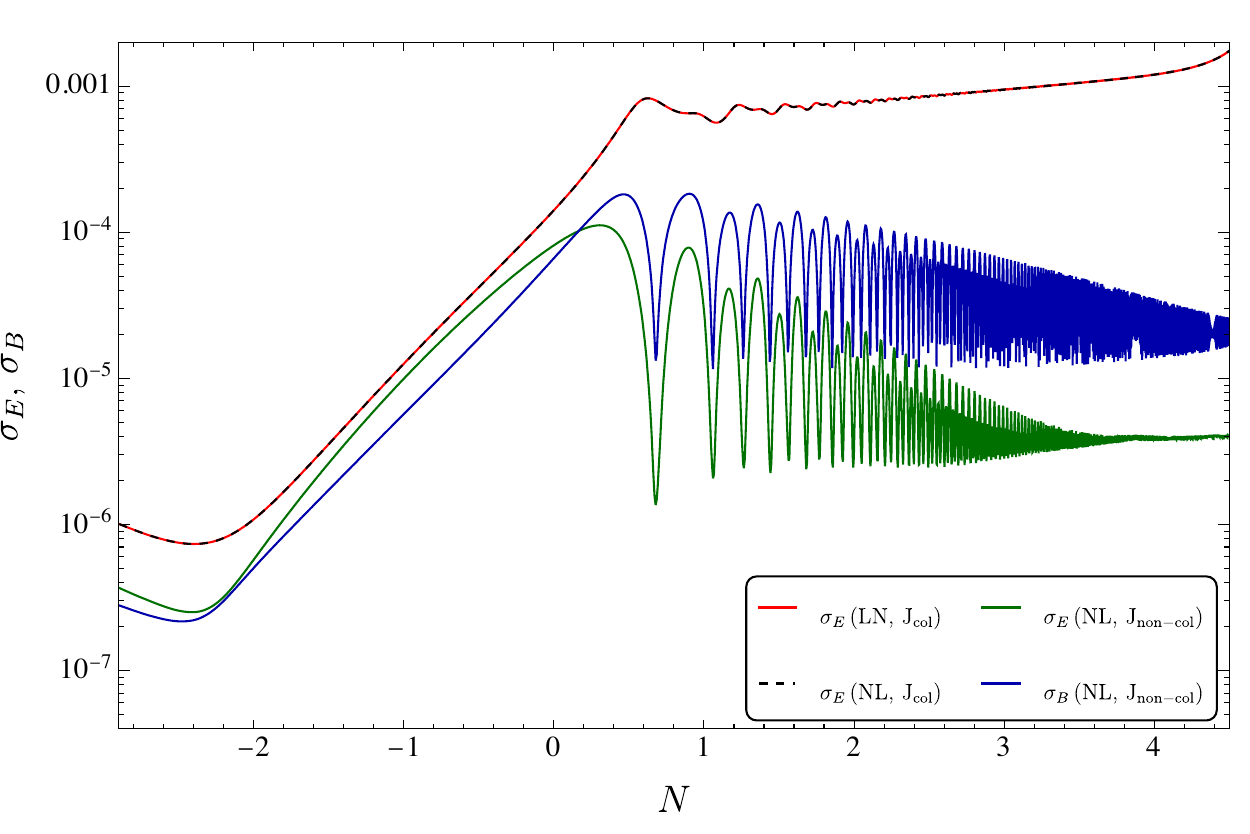}
\end{center}\caption[]{
{\it Upper panel}: The rms values of the comoving electric and magnetic fields under different assumptions.
Magnetic fields are shown for the following cases: linear approximation with a homogeneous inflaton and no Schwinger current (red solid), nonlinear backreaction with a nonhomogeneous inflaton and no Schwinger current (black solid), collinear conductivity description in the linear (orange) and nonlinear (black dashed) regimes, and non-collinear conductivity (blue solid). Finally, the brown-dashed and green-solid curves are the rms electric fields that correspond to the black-dashed and blue-solid $B_{\rm {rms}}$ field results (collinear and non-collinear prescriptions, respectively).
{\it Lower panel:} Electric conductivity in the collinear description is shown for the linear case (red solid) and the nonlinear backreaction case (black dashed). The blue and green solid lines correspond to the non-collinear conductivity prescription. Color coding matches the upper panel.
}\label{fig:EBsigma}\end{figure}

We perform fully nonlinear simulations for the collinear (Eq.~\eqref{eq:sigmaE}) and non-collinear (Eq.~\eqref{eq:mixed}--\eqref{eq:noncolsigma}) description of conductivities, consistently applying the local formulation in the equations of motion. For comparison, we also consider 
the evolution in the strong backreaction regime without fermions (see, e.g., \cite{Sharma:2024nfu, Adshead:2016iae}).  Furthermore, we provide a comparison to  the linear approximation with  homogeneous inflaton dynamics, where $\bm{E} \cdot \bm{B}$ and $\bm{\nabla} \phi$ are omitted in Eq.~\eqref{eq:phi}. The Hubble scale  in the last case is taken to depend only on the inflaton ${\cal H}^2 = (8\pi/3m_{\rm {Pl}}^2)\,a^2\rho_{\phi}$.
The result of our simulations is shown in Figure \ref{fig:EBsigma}. On the top panel we show the evolution of the root-mean-square (rms) magnetic field strength, defined as 
$B^2_{\rm{rms}}=\int 4\pi k dk \, |B(k)|^2$,
and the rms electric field, $E_{\rm{rms}}$, defined analogously. We provide this comparison for collinear and non-collinear cases.

We see that all cases  operating in the large-coupling regime exhibit very similar suppression, regardless of the Schwinger current formulation and the inclusion of  nonlinearities. 
This indicates that the presence of the Schwinger current prevents the gauge field from entering the strong backreaction regime, leading to a universal suppression. As a result, inhomogeneous inflaton gradient contributions remain suppressed, in contrast to studies neglecting the Schwinger effect, where such contributions become particularly important at the end of inflation \cite{Figueroa:2023oxc, Figueroa:2024rkr, Sharma:2024nfu}. Consequently, Schwinger-induced damping can be estimated within the linear approximation, allowing for the use of semi-analytical methods. 

On the lower panel of Fig.~\ref{fig:EBsigma} we show the evolution of conductivities. We see that taking the collinear prescription, the electric conductivity is the same for the linear and non-linear cases, whereas in the non-collinear prescription the electric and magnetic conductivities are oscillating in time.

\noindent\textbf{Universality of the Schwinger backreaction.}
It has been shown  that the largest amplification of gauge fields during axion inflation  occurs close to the end of inflation \cite{Adshead:2015pva},
since  the tachyonic amplification depends on the axion velocity, which is maximal close to the end of inflation.
The growth of $\bm{E}$ and $\bm{B}$ can be described by a simple exponential growth rate (see \cite{Adshead:2015pva} for a WKB analysis).
Furthermore, both fields are almost equal during this growth.
By examining the equation of motion for the gauge field (Eq.~\ref{eq:Edotconf}) 
we see two competing terms:
$(\alpha/ f) (\partial_\tau \phi) \bm{B}$ supports the tachyonic amplification, whereas the current $\bm{J} = \sigma_E \bm{E}$ opposes it.
Initially the tachyonic amplification term dominates and thus one of the two polarizations undergoes  exponential enhancement.
Since $E\approx B$, we can compare the two competing terms by comparing $ (\alpha/ f) \partial_\tau\phi $ to the conductivity $\sigma_E$.
Using $\partial_\tau\phi =m_{\rm{Pl}} {\cal H}   \sqrt{\epsilon/4\pi} $ we get
$(\alpha/ f) \partial_\tau\phi \sim {\cal O}(100) {\cal H} $, where we took $\epsilon\sim 1$ close to the end of inflation and $\alpha m_{\rm {Pl}}/f\sim 60$--$100$. For ${\cal H}\sim 10^{-5}m_{\rm {Pl}}$ we see that the backreaction from Schwinger pair production occurs at $\sigma_E \sim 10^{-3} m_{\rm {Pl}}$. We can  also estimate the typical value of the electric and magnetic fields. First we observe that $\coth(\pi E/B)\simeq 1$ for $E/ B={\cal O}(1)$. Furthermore $eQ\simeq 1$, leading to $B \sim 6\pi^2 (\alpha m_{\rm {Pl}}/f) {\cal H}^2 \sim 10^{-6} m_{\rm {Pl}}^2$. 

Intriguingly, the above estimates for the conductivity and the value of the EM fields at the onset of Schwinger backreaction are consistently supported by our simulations.
We thus uncover  the existence of a universal behavior for axion inflation magnetogenesis, where the Schwinger effect is significant when gauge fields reach a value of $E,B \gtrsim {\cal O}(10^{-6})\, m_{\rm {Pl}}$ close to the end of inflation.
The Schwinger suppression will be less pronounced (largely irrelevant) for couplings that lead to weaker fields.
\\

\begin{figure}[]\begin{center}
\includegraphics[width=\columnwidth]{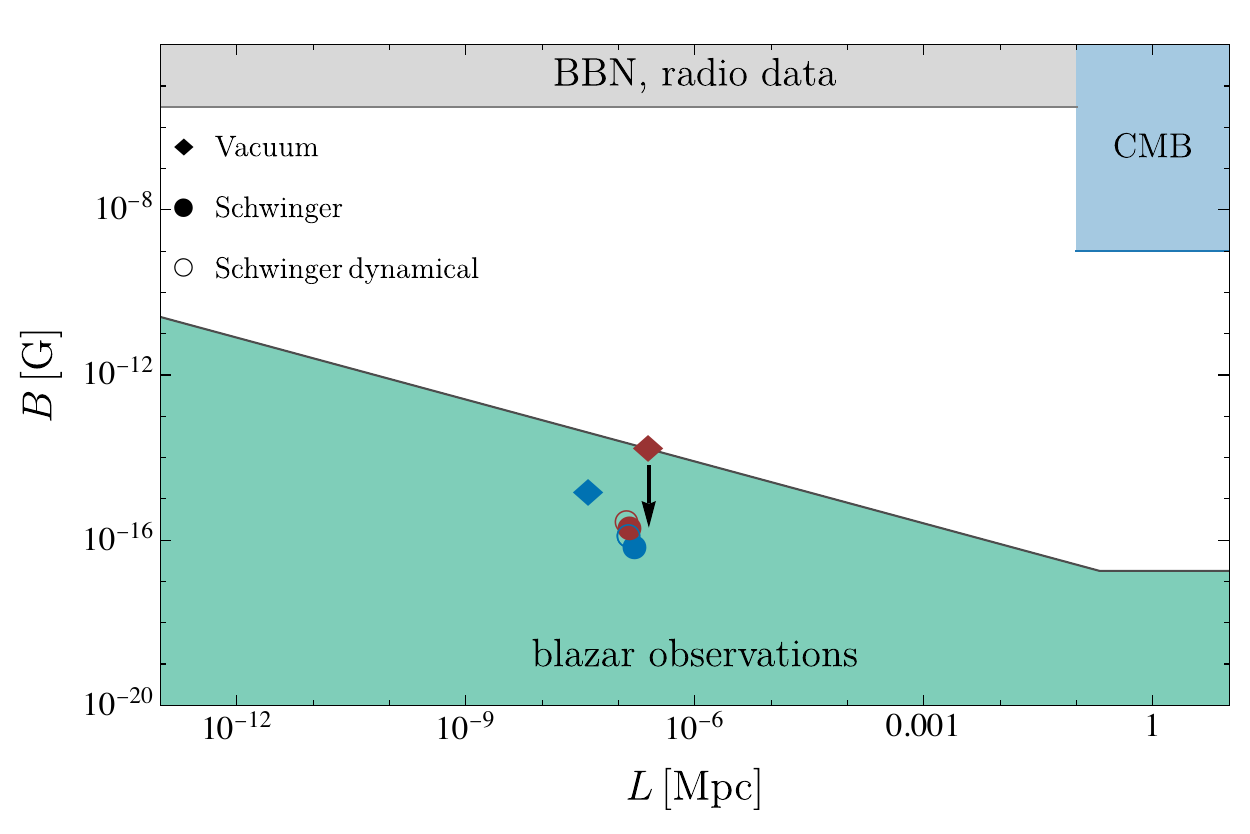}
\end{center}\caption[]{
Magnetic field strength as a function of coherence length for $\alpha m_{\rm {Pl}}/f = 60,90$ (red, blue respectively) without Schwinger current (diamonds), 
and with a presence of a non-dynamical (filled circles) and dynamical (hollow circles) Schwinger current. The blue and gray-shaded areas correspond to excluded upper bounds from Planck \cite{Planck:2015zrl}, Big Bang Nucleosynthesis (BBN) \cite{Kernan:1995bz} and radio data \cite{Kronberg:1993vk}, and the green-shaded area indicates the lower bound on intergalactic magnetic field strength from blazar observations \cite{MAGIC:2022piy}. With the Schwinger current, the values lie below the blazar lower bound. 
}\label{Fig:BLplotmain}\end{figure}

\noindent\textbf{Consequences for magnetogenesis.}
The non-detection of secondary GeV photons from blazars provides indirect evidence for the presence of extragalactic magnetic fields in the intergalactic medium (possibly helical \cite{Vachaspati:2016xji}). This observation motivates investigations into their origin in the early universe. The prospect of magnetogenesis in axion inflation has been explored for couplings up to $\alpha m_{\rm {Pl}}/f \leq 60$~\cite{Adshead:2016iae}  and, more recently, for $\alpha m_{\rm {Pl}}/f = 75, \,90$~\cite{Sharma:2024nfu}. These studies conclude that, for $\alpha m_{\rm {Pl}}/f \geq 60$, the axion–U(1) inflation model is already marginally compatible with generating magnetic fields strong enough to account for the non-observation of GeV photons in blazar spectra. Moreover, as we have seen, the Schwinger effect significantly reduces the final amplitude. To quantify the suppression, let us consider the present-day magnetic field strength and its coherence length, $L$. 
Assuming that the inverse cascade process occurs almost immediately after inflation due to the existence of the Schwinger plasma (see Supplemental Material for details \footnote{See Supplemental Material on the full integration of the current, which includes
Refs.~\cite{
Coleman:1975pw, Dunne:2004nc, Brandenburg:2025ccv, Berger:1984izr, Berger:1984, BP23, BEO96, Kahniashvili+17, Banerjee:2004df, Campanelli:2007tc}.}), the comoving magnetic field strength and correlation length follow $B^2 L={\rm {const}}$.
Figure~\ref{Fig:BLplotmain} shows that the inclusion of the Schwinger effect makes the resulting magnetic field largely independent of the coupling $\alpha/f\ge 60 m_{\rm{Pl}}$ and the prescription for the current (dynamical versus non-dynamical) \footnotemark[2] and well below the lower bound on intergalactic magnetic field strength from blazar observations.

\begin{figure}[t!]\begin{center}
\includegraphics[width=\columnwidth]{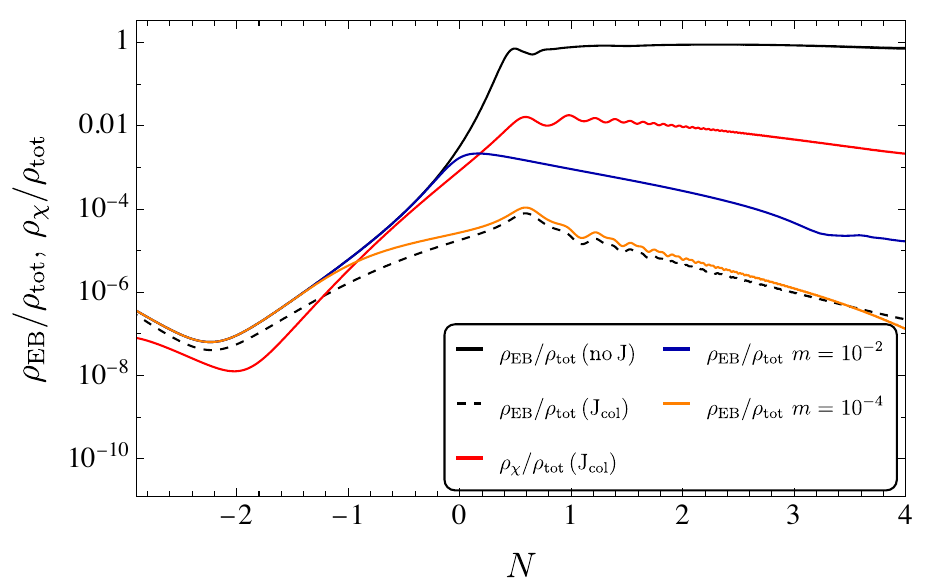}
\end{center}\caption[]{The ratio of the EM energy density $\rho_{EB}\equiv\rho_{E}+\rho_B$ and the plasma energy density $\rho_{\chi}$ to the total energy density $\rho_{\rm{tot}}$ in the simulation box for $\alpha m_{\mathrm{Pl}}/f = 60 $. The black solid curve corresponds to a case without the Schwinger effect and the black-dashed curve to a case with the Schwinger effect and massless fermions. The corresponding plasma energy density ratio is shown in red-solid. The EM energy density ratio for fermion masses $m=10^{-4}, 10^{-2}\, m_{\mathrm{Pl}}$ are shown in solid orange and blue, respectively. 
}\label{fig:rho}\end{figure}

\noindent\textbf{Heavy fermion effects.}
So far, we have neglected the effects arising from finite fermion masses. From \eqref{eq:current} it follows that a significant suppression of the Schwinger effect requires ${\pi m^2 a^2}>{e |Q|E}$ or $m^2 / E_{\rm{ph}}\gtrsim {\cal O}(1)$, where $E_{\rm{ph}}$ is the amplitude of the physical electric field.  We are focusing on the era close to the end of  inflation, where
 $B_{\rm{ph}}^2\simeq E_{\rm{ph}}^2\sim 3H^2 m_{\rm {Pl}}^2/8\pi$, or $E_{\rm{ph}}\sim 0.1 H m_{\rm {Pl}}$. The masses of Standard Model (SM) fermions are given by $m = y h$, where $y$ is the Yukawa coupling and $h$ is the Higgs VEV, which is expected to be nonzero during inflation, if the Higgs is a light field subject to de-Sitter fluctuations. Among the electrically charged fermions, the electron has the smallest Yukawa coupling, $y_e \simeq 3 \times 10^{-6}$, making it the lightest. Thus, if electrons are too heavy to be efficiently produced via the Schwinger effect, all other charged fermions will be even more suppressed.
To suppress the Schwinger effect, the fermion mass must satisfy 
$	m_e^2 \gtrsim 0.1 H m_{\rm {Pl}} 
$.
Figure~\ref{fig:rho} shows how the presence of a large electron mass suppresses the Schwinger current and allows for the generation of a larger magnetic field.
However, the constraint
on $m_s$ implies a lower bound on the Higgs VEV 
$h \gtrsim (0.3/y_e) {\sqrt{H m_{\rm{Pl}}}}  \simeq 10^5 m_{\rm{Pl}}  \sqrt{{H}/{m_{\rm{Pl}}}}$.
To avoid super-Planckian values for the Higgs field, one requires a low inflationary scale: $H \lesssim 10^{-10} m_{\rm{Pl}} \sim 10^9\, \mathrm{GeV}$. This provides a key result: a suppression of the Schwinger effect through fermion masses requires both low-scale inflation and large (possibly Planckian) field excursions for the Higgs.
If one relies solely on de Sitter fluctuations to generate a Higgs VEV, one expects $h \sim H \lambda^{-1/4}$, which leads to $H/m_{\rm {Pl}} \gtrsim 10^{12} \sqrt{\lambda}$, where $\lambda$ is the Higgs self-coupling constant.
This requirement is difficult to satisfy given observational upper bounds on $H$, unless we consider an almost vanishing Higgs self-coupling. However, alternative mechanisms such as a direct coupling between the Higgs and the inflaton, or a non-minimal coupling to gravity can dynamically induce a large Higgs VEV during inflation. 

These considerations point to an intriguing model-building challenge: any realistic suppression of the Schwinger effect involving SM fermions may lead to observable consequences at collider experiments through a modification of the Higgs sector.
Before concluding, we must note that all simulations presented here refer to high-scale inflation,
$H\sim 10^{-6}m_{\rm {Pl}}$, and thus need to be re-done for different Hubble scales. 
In particular, Figure~\ref{fig:rho} shows that the Schwinger effect is still active if we consider large fermion masses, albeit weaker. That being said, reducing the scale of inflaton allows for a large hierarchy between the Hubble scale and thus the possible electric field values are reduced, whereas the fermions  can be equally heavy, if we take the Higgs VEV close to the Planck mass during inflation. Therefore, a definitive conclusion on the viability of axion inflation magnetogenesis requires a dedicated parameter scan, which we defer  for future work.
\\

\noindent\textbf{Summary and outlook.}
We presented the first lattice simulations of the nonlinear evolution after axion inflation
that self-consistently incorporate Schwinger pair production. Our results demonstrate that the induced Schwinger current provides a robust backreaction that quenches gauge field amplification once a \textit{universal critical conductivity} 
$\sigma_E \sim 10^{-3} m_{\rm {Pl}}$ and \textit{magnetic field strength} $B_{\rm{rms}} \sim 10^{-6} m_{\rm {Pl}}^2$ are reached. These critical values are independent of the specific formulation of the current and mark the onset of plasma domination. We found that the Schwinger effect prevents the gauge field from entering the strong backreaction regime, keeping inflaton gradient contributions subdominant and enabling the use of semi-analytical methods.

The resulting suppression of EM field production
significantly reduces the amplitude of primordial magnetic fields, effectively ruling out high scale axion inflation as a source of intergalactic magnetogenesis. 
Avoiding strong suppression requires fermion masses large enough
to inhibit Schwinger production---necessitating a large Higgs VEV during inflation and favoring low-scale inflation scenarios. These findings motivate future studies on low-scale inflationary models and on connections between the axion and Higgs sectors, with potential implications for collider signatures. A different formulation of the Schwinger current, based on a numerical integration of $\partial_\tau J$ provides a very similar result for the present-day $B$ field strength (see Fig.~\ref{Fig:BLplotmain}), which remains well below the level required to account for the blazar observations.
The generation of inhomogeneities in fermion energy density and the resulting
inhomogeneous bulk motions of the plasma remain suppressed during our simulations as we show in the Supplemental Material. However, addressing the full MHD evolution is an
important task for future studies.
Finally, the prevalence of the Schwinger effect in other areas of physics (e.g.~\cite{tolosa2024analog, dunne2009catalysis}), provides an exciting avenue for applying the physical intuition  gained and the code developed here to study these effects. 
\\

\noindent\textbf{Data availability.}
The source code used for the simulations of this study, the {\sc Pencil Code},
is freely available from Refs.~\cite{PencilCode:2020eyn,PC}.
The simulation setups and the corresponding data
are freely available from Ref.~\cite{DATA}.
\\

\noindent\textbf{Acknowledgements.}
The work of O.I.\ was supported by the European Union's Horizon 2020
research and innovation program under the Marie Skłodowska-Curie grant
agreement No.~101106874, and by VR Starting Grant
2025-04140 of the Swedish Research Council. 
O.I.\ is grateful to Leiden University for hospitality, where parts of this work have been completed.
E.I.S. acknowledges support by
the U.S. Department of Energy, Office of Science, Office
of High Energy Physics program under Award Number
DE-SC0022021.
A.B.\ was supported in part by the European Research Council through the ERC Synergy Grant COSMOMAG under grant No.\ 101224803,
the Swedish Research Council (Vetenskapsr{\aa}det) under grant No.\ 2025-05957,
the National Science Foundation under grant Nos.\ NSF AST-2307698, AST-2408411, and NASA Award 80NSSC22K0825.
We thank the Swedish National Allocations Committee for providing computing resources at the Center for Parallel Computers
at the Royal Institute of Technology in Stockholm and the National Supercomputer Centre (NSC) at Link\"oping. We are grateful to the Bernoulli Center and the program ``Generation, Evolution, and Observations of Cosmological Magnetic Fields" for inspiring and motivating this project. We thank Angelo Caravano, Oleksandr Sobol, and Lorenzo Sorbo for comments and discussions during the Nordita program ``Axions in Stockholm, 2025".
\appendix

\bibliography{refs}

@article{Fujita:2022fwc,
    author = "Fujita, Tomohiro and Kume, Jun'ya and Mukaida, Kyohei and Tada, Yuichiro",
    title = "{Effective treatment of U(1) gauge field and charged particles in axion inflation}",
    eprint = "2204.01180",
    archivePrefix = "arXiv",
    primaryClass = "hep-ph",
    reportNumber = "RESCEU-3/22, KEK-TH-2402",
    doi = "10.1088/1475-7516/2022/09/023",
    journal = "JCAP",
    volume = "09",
    pages = "023",
    year = "2022"
}

@article{Figueroa:2023oxc,
    author = "Figueroa, Daniel G. and Lizarraga, Joanes and Urio, Ander and Urrestilla, Jon",
    title = "{Strong Backreaction Regime in Axion Inflation}",
    eprint = "2303.17436",
    archivePrefix = "arXiv",
    primaryClass = "astro-ph.CO",
    doi = "10.1103/PhysRevLett.131.151003",
    journal = "Phys. Rev. Lett.",
    volume = "131",
    number = "15",
    pages = "151003",
    year = "2023"
}

@article{Sobol:2019xls,
    author = "Sobol, O. O. and Gorbar, E. V. and Vilchinskii, S. I.",
    title = "{Backreaction of electromagnetic fields and the Schwinger effect in pseudoscalar inflation magnetogenesis}",
    eprint = "1907.10443",
    archivePrefix = "arXiv",
    primaryClass = "astro-ph.CO",
    doi = "10.1103/PhysRevD.100.063523",
    journal = "Phys. Rev. D",
    volume = "100",
    number = "6",
    pages = "063523",
    year = "2019"
}

@article{Sharma:2024nfu,
    author = "Sharma, Ramkishor and Brandenburg, Axel and Subramanian, Kandaswamy and Vikman, Alexander",
    title = "{Lattice simulations of axion-U(1) inflation: gravitational waves, magnetic fields, and scalar statistics}",
    eprint = "2411.04854",
    archivePrefix = "arXiv",
    primaryClass = "astro-ph.CO",
    reportNumber = "NORDITA-2024-040",
    doi = "10.1088/1475-7516/2025/05/079",
    journal = "JCAP",
    volume = "05",
    pages = "079",
    year = "2025"
}

@ARTICLE{BP23,
       author = {{Brandenburg}, Axel and {Protiti}, Nousaba Nasrin},
        title = "{Electromagnetic Conversion into Kinetic and Thermal Energies}",
      journal = {Entropy},
         year = 2023,
        month = aug,
       volume = {25},
       number = {9},
          eid = {1270},
        pages = {1270},
          doi = {10.3390/e25091270},
archivePrefix = {arXiv},
       eprint = {2308.00662},
 primaryClass = {physics.plasm-ph},
       adsurl = {https://ui.adsabs.harvard.edu/abs/2023Entrp..25.1270B}
}

@article{Bavarsad:2017oyv,
    author = "Bavarsad, Ehsan and Kim, Sang Pyo and Stahl, Cl\'ement and Xue, She-Sheng",
    title = "{Effect of a magnetic field on Schwinger mechanism in de Sitter spacetime}",
    eprint = "1707.03975",
    archivePrefix = "arXiv",
    primaryClass = "hep-th",
    doi = "10.1103/PhysRevD.97.025017",
    journal = "Phys. Rev. D",
    volume = "97",
    number = "2",
    pages = "025017",
    year = "2018"
}

@article{Domcke:2018eki,
    author = "Domcke, Valerie and Mukaida, Kyohei",
    title = "{Gauge Field and Fermion Production during Axion Inflation}",
    eprint = "1806.08769",
    archivePrefix = "arXiv",
    primaryClass = "hep-ph",
    reportNumber = "DESY 18-098, DESY-18-098",
    doi = "10.1088/1475-7516/2018/11/020",
    journal = "JCAP",
    volume = "11",
    pages = "020",
    year = "2018"
}

@article{Domcke:2019qmm,
    author = "Domcke, Valerie and Ema, Yohei and Mukaida, Kyohei",
    title = "{Chiral Anomaly, Schwinger Effect, Euler-Heisenberg Lagrangian, and application to axion inflation}",
    eprint = "1910.01205",
    archivePrefix = "arXiv",
    primaryClass = "hep-ph",
    reportNumber = "DESY-19-166, DESY 19-166",
    doi = "10.1007/JHEP02(2020)055",
    journal = "JHEP",
    volume = "02",
    pages = "055",
    year = "2020"
}

@article{PencilCode:2020eyn,
       author = {{Pencil Code Collaboration} and {Brandenburg}, Axel and {Johansen}, Anders and {Bourdin}, Philippe and {Dobler}, Wolfgang and {Lyra}, Wladimir and {Rheinhardt}, Matthias and {Bingert}, Sven and {Haugen}, Nils and {Mee}, Antony and {Gent}, Frederick and {Babkovskaia}, Natalia and {Yang}, Chao-Chin and {Heinemann}, Tobias and {Dintrans}, Boris and {Mitra}, Dhrubaditya and {Candelaresi}, Simon and {Warnecke}, J{\"o}rn and {K{\"a}pyl{\"a}}, Petri and {Schreiber}, Andreas and {Chatterjee}, Piyali and {K{\"a}pyl{\"a}}, Maarit and {Li}, Xiang-Yu and {Kr{\"u}ger}, Jonas and {Aarnes}, J{\o}rgen and {Sarson}, Graeme and {Oishi}, Jeffrey and {Schober}, Jennifer and {Plasson}, Rapha{\"e}l and {Sandin}, Christer and {Karchniwy}, Ewa and {Rodrigues}, Luiz and {Hubbard}, Alexander and {Guerrero}, Gustavo and {Snodin}, Andrew and {Losada}, Illa and {Pekkil{\"a}}, Johannes and {Qian}, Chengeng},
    collaboration = "Pencil Code",
    title = "{The Pencil Code, a modular MPI code for partial differential equations and particles: multipurpose and multiuser-maintained}",
    eprint = "2009.08231",
    archivePrefix = "arXiv",
    primaryClass = "astro-ph.IM",
    reportNumber = "NORDITA-2020-087",
    doi = "10.21105/joss.02807",
    journal = "J. Open Source Softw.",
    volume = "6",
    number = "58",
    pages = "2807",
    year = "2021"
}

@article{Guth:1980zm,
  author = "Guth, Alan H.",
  title = "{The Inflationary Universe: A Possible Solution to the Horizon and Flatness Problems}",
  journal = "Phys. Rev. D",
  volume = "23",
  year = "1981",
  pages = "347--356",
  doi = "10.1103/PhysRevD.23.347"
}

@article{Linde:1981mu,
  author = "Linde, A. D.",
  title = "{A New Inflationary Universe Scenario: A Possible Solution of the Horizon, Flatness, Homogeneity, Isotropy and Primordial Monopole Problems}",
  journal = "Phys. Lett. B",
  volume = "108",
  year = "1982",
  pages = "389--393",
  doi = "10.1016/0370-2693(82)91219-9"
}

@article{Albrecht:1982wi,
  author = "Albrecht, Andreas and Steinhardt, Paul J.",
  title = "{Cosmology for Grand Unified Theories with Radiatively Induced Symmetry Breaking}",
  journal = "Phys. Rev. Lett.",
  volume = "48",
  year = "1982",
  pages = "1220--1223",
  doi = "10.1103/PhysRevLett.48.1220"
}

@article{Freese:1990rb,
  author = "Freese, Katherine and Frieman, Joshua A. and Olinto, Angela V.",
  title = "{Natural inflation with pseudo - Nambu-Goldstone bosons}",
  journal = "Phys. Rev. Lett.",
  volume = "65",
  year = "1990",
  pages = "3233--3236",
  doi = "10.1103/PhysRevLett.65.3233"
}

@article{Svrcek:2006yi,
  author = "Svrcek, Peter and Witten, Edward",
  title = "{Axions In String Theory}",
  journal = "JHEP",
  volume = "06",
  year = "2006",
  pages = "051",
  doi = "10.1088/1126-6708/2006/06/051",
  eprint = "hep-th/0605206"
}

@article{Anber:2009ua,
  author = "Anber, Mohamed M. and Sorbo, Lorenzo",
  title = "{Naturally inflating on steep potentials through electromagnetic dissipation}",
  journal = "Phys. Rev. D",
  volume = "81",
  year = "2010",
  pages = "043534",
  doi = "10.1103/PhysRevD.81.043534",
  eprint = "0908.4089"
}

@article{Barnaby:2011qe,
  author = "Barnaby, Neil and Namba, Ryo and Peloso, Marco",
  title = "{Phenomenology of a pseudo-scalar inflaton: Naturally large non-Gaussianity}",
  journal = "JCAP",
  volume = "04",
  year = "2011",
  pages = "009",
  doi = "10.1088/1475-7516/2011/04/009",
  eprint = "1102.4333"
}

@article{Barnaby:2010vf,
  author = "Barnaby, Neil and Peloso, Marco",
  title = "{Large non-Gaussianity in axion inflation}",
  journal = "Phys. Rev. Lett.",
  volume = "106",
  year = "2011",
  pages = "181301",
  doi = "10.1103/PhysRevLett.106.181301",
  eprint = "1011.1500"
}

@article{Kluger:1991ib,
  author = "Kluger, Y. and Eisenberg, J. M. and Svetitsky, B. and Cooper, Fred and Mottola, Emil",
  title = "{Fermion pair production in a strong electric field}",
  journal = "Phys. Rev. D",
  volume = "45",
  year = "1992",
  pages = "4659--4671",
  doi = "10.1103/PhysRevD.45.4659"
}

@article{Adshead:2015pva,
    author = "Adshead, Peter and Giblin, John T. and Scully, Timothy R. and Sfakianakis, Evangelos I.",
    title = "{Gauge-preheating and the end of axion inflation}",
    eprint = "1502.06506",
    archivePrefix = "arXiv",
    primaryClass = "astro-ph.CO",
    doi = "10.1088/1475-7516/2015/12/034",
    journal = "JCAP",
    volume = "12",
    pages = "034",
    year = "2015"
}

@article{Adshead:2016iae,
    author = "Adshead, Peter and Giblin, John T. and Scully, Timothy R. and Sfakianakis, Evangelos I.",
    title = "{Magnetogenesis from axion inflation}",
    eprint = "1606.08474",
    archivePrefix = "arXiv",
    primaryClass = "astro-ph.CO",
    doi = "10.1088/1475-7516/2016/10/039",
    journal = "JCAP",
    volume = "10",
    pages = "039",
    year = "2016"
}

@article{MAGIC:2022piy,
    author = "Acciari, V. A. and others",
    collaboration = "MAGIC",
    title = "{A lower bound on intergalactic magnetic fields from time variability of 1ES 0229+200 from MAGIC and Fermi/LAT observations}",
    eprint = "2210.03321",
    archivePrefix = "arXiv",
    primaryClass = "astro-ph.HE",
    doi = "10.1051/0004-6361/202244126",
    journal = "Astron. Astrophys.",
    volume = "670",
    pages = "A145",
    year = "2023"
}

@article{Vachaspati:2016xji,
    author = "Vachaspati, Tanmay",
    title = "{Fundamental Implications of Intergalactic Magnetic Field Observations}",
    eprint = "1606.06186",
    archivePrefix = "arXiv",
    primaryClass = "astro-ph.CO",
    doi = "10.1103/PhysRevD.95.063505",
    journal = "Phys. Rev. D",
    volume = "95",
    number = "6",
    pages = "063505",
    year = "2017"
}

@article{Barnaby:2011vw,
    author = "Barnaby, Neil and Namba, Ryo and Peloso, Marco",
    title = "{Phenomenology of a Pseudo-Scalar Inflaton: Naturally Large Nongaussianity}",
    eprint = "1102.4333",
    archivePrefix = "arXiv",
    primaryClass = "astro-ph.CO",
    doi = "10.1088/1475-7516/2011/04/009",
    journal = "JCAP",
    volume = "04",
    pages = "009",
    year = "2011"
}

@article{Sorbo:2011rz,
    author = "Sorbo, Lorenzo",
    title = "{Parity violation in the Cosmic Microwave Background from a pseudoscalar inflaton}",
    eprint = "1101.1525",
    archivePrefix = "arXiv",
    primaryClass = "astro-ph.CO",
    doi = "10.1088/1475-7516/2011/06/003",
    journal = "JCAP",
    volume = "06",
    pages = "003",
    year = "2011"
}

@article{Cook:2013xea,
    author = "Cook, Jessica L. and Sorbo, Lorenzo",
    title = "{An inflationary model with small scalar and large tensor nongaussianities}",
    eprint = "1307.7077",
    archivePrefix = "arXiv",
    primaryClass = "astro-ph.CO",
    doi = "10.1088/1475-7516/2013/11/047",
    journal = "JCAP",
    volume = "11",
    pages = "047",
    year = "2013"
}

@article{Bastero-Gil:2022fme,
    author = "Bastero-Gil, Mar and Manso, Ant\'onio Torres",
    title = "{Parity violating gravitational waves at the end of inflation}",
    eprint = "2209.15572",
    archivePrefix = "arXiv",
    primaryClass = "gr-qc",
    doi = "10.1088/1475-7516/2023/08/001",
    journal = "JCAP",
    volume = "08",
    pages = "001",
    year = "2023"
}

@article{Garcia-Bellido:2023ser,
    author = "Garcia-Bellido, Juan and Papageorgiou, Alexandros and Peloso, Marco and Sorbo, Lorenzo",
    title = "{A flashing beacon in axion inflation: recurring bursts of gravitational waves in the strong backreaction regime}",
    eprint = "2303.13425",
    archivePrefix = "arXiv",
    primaryClass = "astro-ph.CO",
    doi = "10.1088/1475-7516/2024/01/034",
    journal = "JCAP",
    volume = "01",
    pages = "034",
    year = "2024"
}

@article{Garretson:1992vt,
    author = "Garretson, W. Daniel and Field, George B. and Carroll, Sean M.",
    title = "{Primordial magnetic fields from pseudoGoldstone bosons}",
    eprint = "hep-ph/9209238",
    archivePrefix = "arXiv",
    reportNumber = "PRINT-92-0448 (CFA,CAMBRIDGE), CFA-3507",
    doi = "10.1103/PhysRevD.46.5346",
    journal = "Phys. Rev. D",
    volume = "46",
    pages = "5346--5351",
    year = "1992"
}

@article{Anber:2006xt,
    author = "Anber, Mohamed M. and Sorbo, Lorenzo",
    title = "{N-flationary magnetic fields}",
    eprint = "astro-ph/0606534",
    archivePrefix = "arXiv",
    doi = "10.1088/1475-7516/2006/10/018",
    journal = "JCAP",
    volume = "10",
    pages = "018",
    year = "2006"
}

@article{Durrer:2023rhc,
    author = "Durrer, R. and Sobol, O. and Vilchinskii, S.",
    title = "{Backreaction from gauge fields produced during inflation}",
    eprint = "2303.04583",
    archivePrefix = "arXiv",
    primaryClass = "gr-qc",
    reportNumber = "MS-TP-23-06",
    doi = "10.1103/PhysRevD.108.043540",
    journal = "Phys. Rev. D",
    volume = "108",
    number = "4",
    pages = "043540",
    year = "2023"
}

@article{Cuissa:2018oiw,
    author = "Cuissa, Jose Roberto Canivete and Figueroa, Daniel G.",
    title = "{Lattice formulation of axion inflation. Application to preheating}",
    eprint = "1812.03132",
    archivePrefix = "arXiv",
    primaryClass = "astro-ph.CO",
    doi = "10.1088/1475-7516/2019/06/002",
    journal = "JCAP",
    volume = "06",
    pages = "002",
    year = "2019"
}

@article{Adshead:2023mvt,
    author = "Adshead, Peter and Giblin, John T. and Grutkoski, Ryn and Weiner, Zachary J.",
    title = "{Gauge preheating with full general relativity}",
    eprint = "2311.01504",
    archivePrefix = "arXiv",
    primaryClass = "astro-ph.CO",
    doi = "10.1088/1475-7516/2024/03/017",
    journal = "JCAP",
    volume = "03",
    pages = "017",
    year = "2024"
}

@article{Adshead:2019igv, 
    author = "Adshead, Peter and Giblin, John T. and Pieroni, Mauro and Weiner, Zachary J.",
    title = "{Constraining Axion Inflation with Gravitational Waves across 29 Decades in Frequency}",
    eprint = "1909.12843",
    archivePrefix = "arXiv",
    primaryClass = "astro-ph.CO",
    doi = "10.1103/PhysRevLett.124.171301",
    journal = "Phys. Rev. Lett.",
    volume = "124",
    number = "17",
    pages = "171301",
    year = "2020"
}

@article{Adshead:2019lbr,
    author = "Adshead, Peter and Giblin, John T. and Pieroni, Mauro and Weiner, Zachary J.",
    title = "{Constraining axion inflation with gravitational waves from preheating}",
    eprint = "1909.12842",
    archivePrefix = "arXiv",
    primaryClass = "astro-ph.CO",
    doi = "10.1103/PhysRevD.101.083534",
    journal = "Phys. Rev. D",
    volume = "101",
    number = "8",
    pages = "083534",
    year = "2020"
}

@article{Caravano:2021bfn,
    author = "Caravano, Angelo and Komatsu, Eiichiro and Lozanov, Kaloian D. and Weller, Jochen",
    title = "{Lattice simulations of Abelian gauge fields coupled to axions during inflation}",
    eprint = "2110.10695",
    archivePrefix = "arXiv",
    primaryClass = "astro-ph.CO",
    doi = "10.1103/PhysRevD.105.123530",
    journal = "Phys. Rev. D",
    volume = "105",
    number = "12",
    pages = "123530",
    year = "2022"
}

@article{Caravano:2022epk,
    author = "Caravano, Angelo and Komatsu, Eiichiro and Lozanov, Kaloian D. and Weller, Jochen",
    title = "{Lattice simulations of axion-U(1) inflation}",
    eprint = "2204.12874",
    archivePrefix = "arXiv",
    primaryClass = "astro-ph.CO",
    doi = "10.1103/PhysRevD.108.043504",
    journal = "Phys. Rev. D",
    volume = "108",
    number = "4",
    pages = "043504",
    year = "2023"
}

@article{Caravano:2024xsb,
    author = "Caravano, Angelo and Peloso, Marco",
    title = "{Unveiling the nonlinear dynamics of a rolling axion during inflation}",
    eprint = "2407.13405",
    archivePrefix = "arXiv",
    primaryClass = "astro-ph.CO",
    doi = "10.1088/1475-7516/2025/01/104",
    journal = "JCAP",
    volume = "01",
    pages = "104",
    year = "2025"
}

@article{Figueroa:2024rkr,
    author = "Figueroa, Daniel G. and Lizarraga, Joanes and Loayza, Nicol\'as and Urio, Ander and Urrestilla, Jon",
    title = "{Nonlinear dynamics of axion inflation: A detailed lattice study}",
    eprint = "2411.16368",
    archivePrefix = "arXiv",
    primaryClass = "astro-ph.CO",
    doi = "10.1103/PhysRevD.111.063545",
    journal = "Phys. Rev. D",
    volume = "111",
    number = "6",
    pages = "063545",
    year = "2025"
}

@article{Sauter:1931zz,
    author = "Sauter, Fritz",
    title = "{Uber das Verhalten eines Elektrons im homogenen elektrischen Feld nach der relativistischen Theorie Diracs}",
    doi = "10.1007/BF01339461",
    journal = "Z. Phys.",
    volume = "69",
    pages = "742--764",
    year = "1931"
}

@article{Heisenberg:1936nmg,
    author = "Heisenberg, W. and Euler, H.",
    title = "{Consequences of Dirac's theory of positrons}",
    eprint = "physics/0605038",
    archivePrefix = "arXiv",
    doi = "10.1007/BF01343663",
    journal = "Z. Phys.",
    volume = "98",
    number = "11-12",
    pages = "714--732",
    year = "1936"
}

@article{Fujita:2015iga,
    author = "Fujita, Tomohiro and Namba, Ryo and Tada, Yuichiro and Takeda, Naoyuki and Tashiro, Hiroyuki",
    title = "{Consistent generation of magnetic fields in axion inflation models}",
    eprint = "1503.05802",
    archivePrefix = "arXiv",
    primaryClass = "astro-ph.CO",
    reportNumber = "IPMU-15-0029, ICRR-REPORT-699-2014-25",
    doi = "10.1088/1475-7516/2015/05/054",
    journal = "JCAP",
    volume = "05",
    pages = "054",
    year = "2015"
}

@article{Gorbar:2021rlt,
    author = "Gorbar, E. V. and Schmitz, K. and Sobol, O. O. and Vilchinskii, S. I.",
    title = "{Gauge-field production during axion inflation in the gradient expansion formalism}",
    eprint = "2109.01651",
    archivePrefix = "arXiv",
    primaryClass = "hep-ph",
    reportNumber = "CERN-TH-2021-128",
    doi = "10.1103/PhysRevD.104.123504",
    journal = "Phys. Rev. D",
    volume = "104",
    number = "12",
    pages = "123504",
    year = "2021"
}

@article{Gorbar:2023zla,
    author = "Gorbar, E. V. and Momot, A. I. and Prikhodko, O. O. and Teslyk, O. M.",
    title = "{Hydrodynamical approach to chirality production during axion inflation}",
    eprint = "2311.07429",
    archivePrefix = "arXiv",
    primaryClass = "hep-ph",
    doi = "10.1103/PhysRevD.109.023536",
    journal = "Phys. Rev. D",
    volume = "109",
    number = "2",
    pages = "023536",
    year = "2024"
}

@article{vonEckardstein:2024tix,
    author = "von Eckardstein, Richard and Schmitz, Kai and Sobol, Oleksandr",
    title = "{On the Schwinger effect during axion inflation}",
    eprint = "2408.16538",
    archivePrefix = "arXiv",
    primaryClass = "hep-ph",
    reportNumber = "MS-TP-24-20",
    doi = "10.1007/JHEP02(2025)096",
    journal = "JHEP",
    volume = "02",
    pages = "096",
    year = "2025"
}

@article{Domcke:2016bkh,
    author = "Domcke, Valerie and Pieroni, Mauro and Bin\'etruy, Pierre",
    title = "{Primordial gravitational waves for universality classes of pseudoscalar inflation}",
    eprint = "1603.01287",
    archivePrefix = "arXiv",
    primaryClass = "astro-ph.CO",
    doi = "10.1088/1475-7516/2016/06/031",
    journal = "JCAP",
    volume = "06",
    pages = "031",
    year = "2016"
}

@article{Garcia-Bellido:2016dkw,
    author = "Garcia-Bellido, Juan and Peloso, Marco and Unal, Caner",
    title = "{Gravitational waves at interferometer scales and primordial black holes in axion inflation}",
    eprint = "1610.03763",
    archivePrefix = "arXiv",
    primaryClass = "astro-ph.CO",
    reportNumber = "IFT-UAM-CSIC-16-100, UMN-TH-3607-16",
    doi = "10.1088/1475-7516/2016/12/031",
    journal = "JCAP",
    volume = "12",
    pages = "031",
    year = "2016"
}

@article{Corba:2024tfz,
    author = "Corb\`a, Sofia P. and Sorbo, Lorenzo",
    title = "{Correlated scalar perturbations and gravitational waves from axion inflation}",
    eprint = "2403.03338",
    archivePrefix = "arXiv",
    primaryClass = "astro-ph.CO",
    doi = "10.1088/1475-7516/2024/10/024",
    journal = "JCAP",
    volume = "10",
    pages = "024",
    year = "2024"
}

@article{Banerjee:2004df,
    author = "Banerjee, Robi and Jedamzik, Karsten",
    title = "{The Evolution of cosmic magnetic fields: From the very early universe, to recombination, to the present}",
    eprint = "astro-ph/0410032",
    archivePrefix = "arXiv",
    doi = "10.1103/PhysRevD.70.123003",
    journal = "Phys. Rev. D",
    volume = "70",
    pages = "123003",
    year = "2004"
}

@article{Campanelli:2007tc,
    author = "Campanelli, Leonardo",
    title = "{Evolution of Magnetic Fields in Freely Decaying Magnetohydrodynamic Turbulence}",
    eprint = "0705.2308",
    archivePrefix = "arXiv",
    primaryClass = "astro-ph",
    doi = "10.1103/PhysRevLett.98.251302",
    journal = "Phys. Rev. Lett.",
    volume = "98",
    pages = "251302",
    year = "2007"
}

@article{Baumann:2009ni,
    author = "Baumann, Daniel and McAllister, Liam",
    title = "{Advances in Inflation in String Theory}",
    eprint = "0901.0265",
    archivePrefix = "arXiv",
    primaryClass = "hep-th",
    doi = "10.1146/annurev.nucl.010909.083524",
    journal = "Ann. Rev. Nucl. Part. Sci.",
    volume = "59",
    pages = "67--94",
    year = "2009"
}

@article{Blumenhagen:2014gta,
    author = "Blumenhagen, Ralph and Plauschinn, Erik",
    title = "{Towards Universal Axion Inflation and Reheating in String Theory}",
    eprint = "1404.3542",
    archivePrefix = "arXiv",
    primaryClass = "hep-th",
    doi = "10.1016/j.physletb.2014.08.007",
    journal = "Phys. Lett. B",
    volume = "736",
    pages = "482--487",
    year = "2014"
}

@article{Palti:2015xra,
    author = "Palti, Eran",
    title = "{On Natural Inflation and Moduli Stabilisation in String Theory}",
    eprint = "1508.00009",
    archivePrefix = "arXiv",
    primaryClass = "hep-th",
    doi = "10.1007/JHEP10(2015)188",
    journal = "JHEP",
    volume = "10",
    pages = "188",
    year = "2015"
}

@article{Czerny:2014xja,
    author = "Czerny, Michael and Higaki, Tetsutaro and Takahashi, Fuminobu",
    title = "{Multi-Natural Inflation in Supergravity}",
    eprint = "1403.0410",
    archivePrefix = "arXiv",
    primaryClass = "hep-ph",
    reportNumber = "KEK-TH-1710, TU-956, IPMU14-0048",
    doi = "10.1007/JHEP05(2014)144",
    journal = "JHEP",
    volume = "05",
    pages = "144",
    year = "2014"
}

@article{Higaki:2014pja,
    author = "Higaki, Tetsutaro and Takahashi, Fuminobu",
    title = "{Natural and Multi-Natural Inflation in Axion Landscape}",
    eprint = "1404.6923",
    archivePrefix = "arXiv",
    primaryClass = "hep-th",
    reportNumber = "KEK-TH-1730, IPMU14-0107, TU-967",
    doi = "10.1007/JHEP07(2014)074",
    journal = "JHEP",
    volume = "07",
    pages = "074",
    year = "2014"
}

@ARTICLE{DATA,
       author = {{Iarygina}, O. and {Sfakianakis}, E. I. and {Brandenburg}, A.},
      journal = "{Datasets for Schwinger effect in axion inflation on a lattice; see \url{http://norlx65.nordita.org/~brandenb/projects/Schwinger}}"
}

@misc{PC,
        title = {The Pencil Code.  DOI:10.5281/zenodo.2315093.  \href{https://github.com/pencil-code}{https://github.com/pencil-code}}
}

@article{McAllister:2008hb,
    author = "McAllister, Liam and Silverstein, Eva and Westphal, Alexander",
    title = "{Gravity Waves and Linear Inflation from Axion Monodromy}",
    eprint = "0808.0706",
    archivePrefix = "arXiv",
    primaryClass = "hep-th",
    reportNumber = "SLAC-PUB-13357, SU-ITP-08-15",
    doi = "10.1103/PhysRevD.82.046003",
    journal = "Phys. Rev. D",
    volume = "82",
    pages = "046003",
    year = "2010"
}

@article{Schwinger:1951nm,
    author = "Schwinger, Julian S.",
    editor = "Milton, K. A.",
    title = "{On gauge invariance and vacuum polarization}",
    doi = "10.1103/PhysRev.82.664",
    journal = "Phys. Rev.",
    volume = "82",
    pages = "664--679",
    year = "1951"
}

@article{Coleman:1975pw,
    author = "Coleman, Sidney R. and Jackiw, R. and Susskind, Leonard",
    title = "{Charge Shielding and Quark Confinement in the Massive Schwinger Model}",
    reportNumber = "MIT-CTP-470",
    doi = "10.1016/0003-4916(75)90212-2",
    journal = "Annals Phys.",
    volume = "93",
    pages = "267",
    year = "1975"
}

@inbook{Dunne:2004nc,
    author = "Dunne, Gerald V.",
    editor = "Shifman, M. and Vainshtein, A. and Wheater, J.",
    title = "{Heisenberg-Euler effective Lagrangians: Basics and extensions}",
    booktitle = "{From fields to strings: Circumnavigating theoretical physics. Ian Kogan memorial collection (3 volume set)}",
    eprint = "hep-th/0406216",
    archivePrefix = "arXiv",
    doi = "10.1142/9789812775344_0014",
    pages = "445--522",
    month = "6",
    year = "2004"
}

@ARTICLE{BEO96,
       author = {{Brandenburg}, Axel and {Enqvist}, Kari and {Olesen}, Poul},
        title = "{Large-scale magnetic fields from hydromagnetic turbulence in the very early universe}",
      journal = {\prd},
         year = 1996,
        month = jul,
       volume = {54},
       number = {2},
        pages = {1291-1300},
          doi = {10.1103/PhysRevD.54.1291},
archivePrefix = {arXiv},
       eprint = {astro-ph/9602031},
 primaryClass = {astro-ph},
       adsurl = {https://ui.adsabs.harvard.edu/abs/1996PhRvD..54.1291B}
}

@ARTICLE{Kahniashvili+17,
       author = {{Kahniashvili}, Tina and {Brandenburg}, Axel and {Durrer}, Ruth and {Tevzadze}, Alexander G. and {Yin}, Winston},
        title = "{Scale-invariant helical magnetic field evolution and the duration of inflation}",
      journal = {JCAP},
         year = 2017,
        month = dec,
       volume = {2017},
       number = {12},
          eid = {002},
        pages = {002},
          doi = {10.1088/1475-7516/2017/12/002},
archivePrefix = {arXiv},
       eprint = {1610.03139},
 primaryClass = {astro-ph.CO},
       adsurl = {https://ui.adsabs.harvard.edu/abs/2017JCAP...12..002K}
}

@article{dunne2009catalysis,
  title = {Catalysis of Schwinger Vacuum Pair Production},
  author = {Dunne, Gerald V. and Gies, Holger and Schützhold, Ralf},
  journal = {Physical Review D},
  volume = {80},
  number = {11},
  pages = {111301},
  year = {2009},
  doi = {10.1103/PhysRevD.80.111301},
  archivePrefix = {arXiv},
  eprint = {0908.0948}
}

@article{tolosa2024analog,
  title={Analog of cosmological particle production in Dirac materials},
  author={Tolosa-Sime{\'o}n, Mireia and Scherer, Michael M. and Floerchinger, Stefan},
  journal={Physical Review B},
  volume={110},
  number={8},
  pages={085421},
  year={2024},
  doi={10.1103/PhysRevB.110.085421}
}

@article{Planck:2015zrl,
    author = "Ade, P. A. R. and others",
    collaboration = "Planck",
    title = "{Planck 2015 results. XIX. Constraints on primordial magnetic fields}",
    eprint = "1502.01594",
    archivePrefix = "arXiv",
    primaryClass = "astro-ph.CO",
    doi = "10.1051/0004-6361/201525821",
    journal = "Astron. Astrophys.",
    volume = "594",
    pages = "A19",
    year = "2016"
}

@article{Kronberg:1993vk,
    author = "Kronberg, Philipp P.",
    title = "{Extragalactic magnetic fields}",
    doi = "10.1088/0034-4885/57/4/001",
    journal = "Rept. Prog. Phys.",
    volume = "57",
    pages = "325--382",
    year = "1994"
}

@article{Brandenburg:2025ccv,
    author = "Brandenburg, Axel and Banerjee, Aikya",
    title = "{Turbulent magnetic decay controlled by two conserved quantities}",
    doi = "10.1017/s0022377824001508",
    journal = "J. Plasma Phys.",
    volume = "91",
    number = "1",
    pages = "E5",
    year = "2025"
}

@article{Kernan:1995bz,
    author = "Kernan, Peter J. and Starkman, Glenn D. and Vachaspati, Tanmay",
    title = "{Big bang nucleosynthesis constraints on primordial magnetic fields}",
    eprint = "astro-ph/9509126",
    archivePrefix = "arXiv",
    reportNumber = "CWRU-P10-95",
    doi = "10.1103/PhysRevD.54.7207",
    journal = "Phys. Rev. D",
    volume = "54",
    pages = "7207--7214",
    year = "1996"
}

@article{Berger:1984,
    author = "Berger, Mitchell A.",
    title = "{Rigorous new limits on magnetic helicity dissipation in the solar corona}",
    doi = "10.1080/03091928408210078",
    journal = "Geophys. Astrophys. Fluid Dyn.",
    volume = "30",
    pages = "79--104",
    year = "1984"
}

@article{Berger:1984izr,
    author = "Berger, Mitchell A. and Field, George B.",
    title = "{The topological properties of magnetic helicity}",
    doi = "10.1017/s0022112084002019",
    journal = "J. Fluid Mech.",
    volume = "147",
    pages = "133--148",
    year = "1984"
}
 
\clearpage
\newpage

\onecolumngrid
\setcounter{secnumdepth}{3}
\setcounter{equation}{0}
\setcounter{figure}{0}
\setcounter{table}{0}
\setcounter{page}{1}
\makeatletter
\renewcommand{\theequation}{S\arabic{equation}}
\renewcommand{\thefigure}{S\arabic{figure}}
\renewcommand{\bibnumfmt}[1]{[#1]}
\renewcommand{\citenumfont}[1]{#1}
\pagestyle{plain}

\begin{center}
\Large{\textbf{Schwinger effect in axion inflation on a lattice}}\\
\medskip
\textit{Supplemental Material}\\
\medskip
{Oksana Iarygina, Evangelos I. Sfakianakis, and Axel Brandenburg}
\end{center}

\subsection{Evolution with different couplings}
\label{supp:alphas}
\begin{figure}[h!]\begin{center}
\includegraphics[width=0.45\columnwidth]{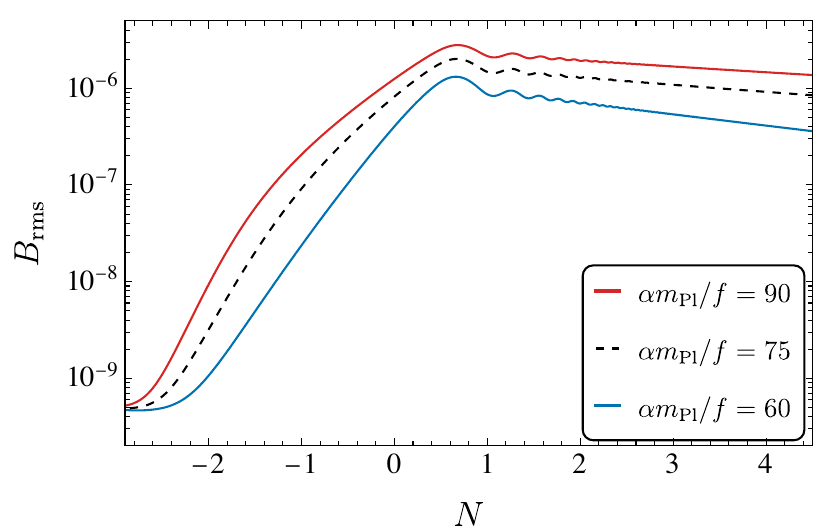}
\includegraphics[width=0.45\columnwidth]{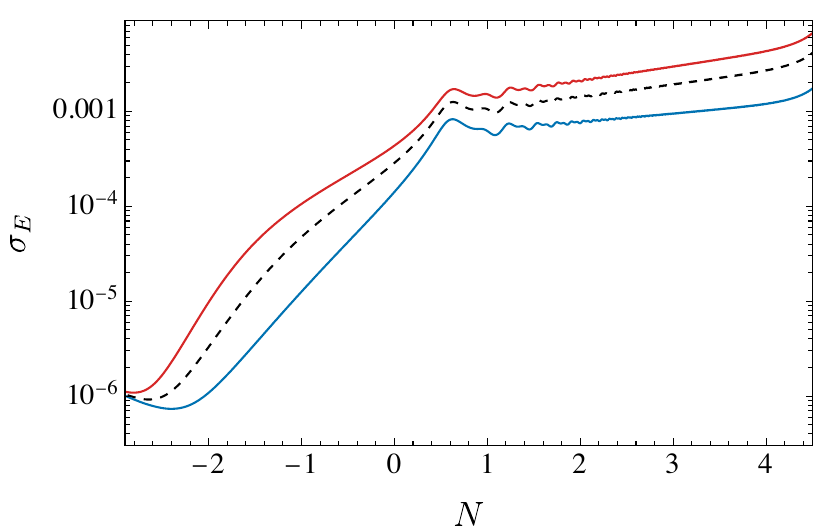}
\end{center}\caption[]{
Comparison of comoving ${B}_\mathrm{rms}$ and $\sigma_E$ 
for cases with $\alpha m_{\rm {Pl}}/f=90 , 75,\, 60$ (solid red, dashed black, solid blue curves) for the collinear case and full nonlinear evolution.  The figure illustrates that Schwinger backreaction universally suppresses the magnetic field, with variations remaining within about an order of magnitude at strong coupling. 
}\label{fig:alphas}\end{figure}

 In this section we demonstrate the evolution of the magnetic field and electric conductivities for coupling constants $\alpha m_{\rm {Pl}}/f=90$ (solid red), $75$ (dashed black) and $60$ (solid blue).
 For simplicity we chose the collinear case for conductivities given by Eq.~\eqref{eq:sigmaE}. As illustrated in Figure \ref{fig:alphas}, the impact of Schwinger suppression remains within an order-of-magnitude variation at strong coupling, highlighting the universality of the Schwinger backreaction. 
 
\subsection{Full integration of the current}
\label{supp:current_time}

Let us start by providing a heuristic derivation of the Schwinger current \cite{Schwinger:1951nm, Coleman:1975pw} in Minkowski space.  We first compute the pair production rate \cite{Dunne:2004nc} (for weak fields) $\Gamma \propto (eE)^2 e^{-\pi m^2\over eE}$. These particles are accelerated  by the electric field through $v = {e\over m}\int E\, dt$. Combining these leads to an equation for the current as an integral over time, or equivalently a first-order differential equation where $dJ/dt$ is a function of $E$. 
Performing a more careful calculation in an expanding space-time similarly leads to the first-order differential equation for the dynamical current
\cite{ Domcke:2018eki}
\begin{equation}
\label{eq:Jderiv}
    \partial_{\tau}J=\frac{(e |Q|)^3}{2\pi^2}E |B| \coth\left( \frac{\pi |B|}{E}\right)
    \, ,
\end{equation}
where  the fermion mass is  dropped for simplicity and the fields are comoving as in the main text.
One can integrate the above, by assuming constant (physical) EM fields and constant Hubble scale, and arrive at
\begin{equation}
\label{eq:Jintegrated}
      J=\frac{(e |Q|)^3}{6\pi^2 {\cal H}}E |B| \coth\left( \frac{\pi |B|}{E}\right)
\end{equation}
which is the form of the current that we used in our simulations and has been widely used in the literature \cite{Domcke:2018eki, Sobol:2019xls, Domcke:2019qmm, Fujita:2022fwc, vonEckardstein:2024tix}.

\begin{figure}[h!]\begin{center}
\includegraphics[width=0.45\columnwidth]{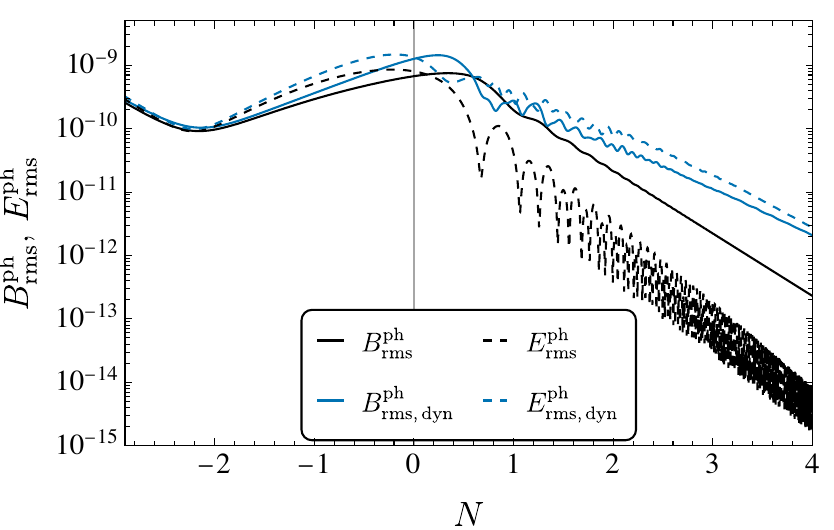}
\includegraphics[width=0.45\columnwidth]{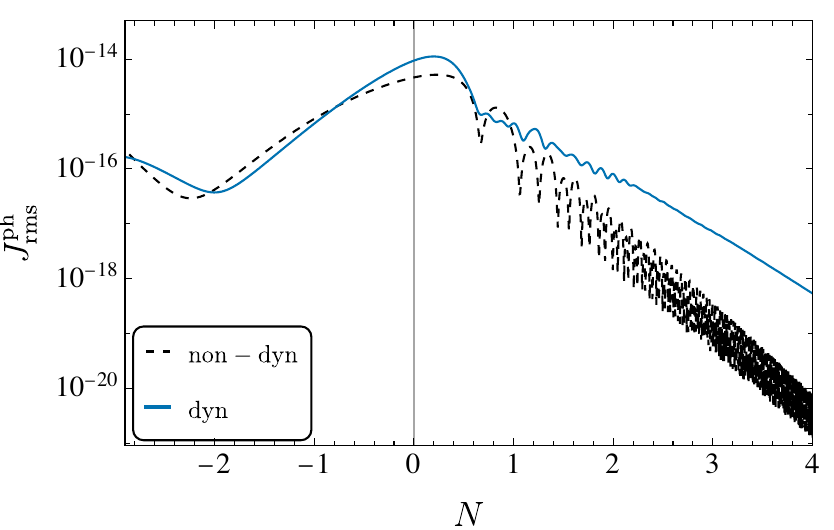}\\
\includegraphics[width=0.45\columnwidth]{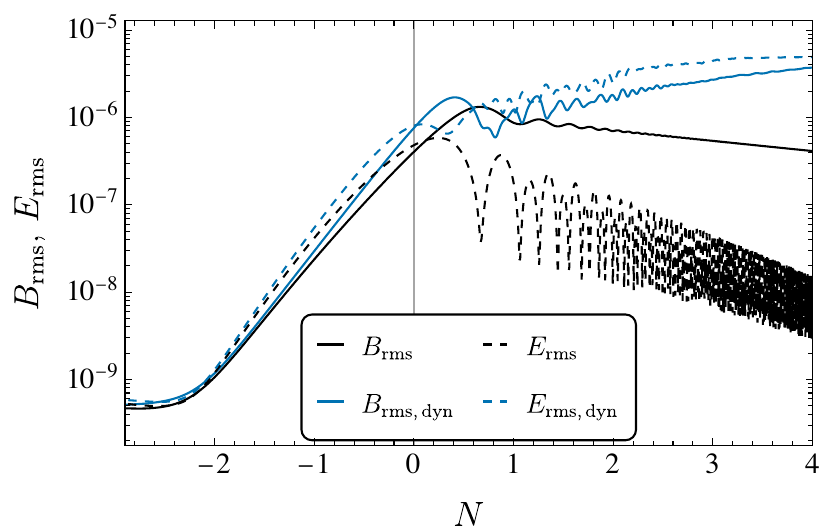}
\includegraphics[width=0.45\columnwidth]{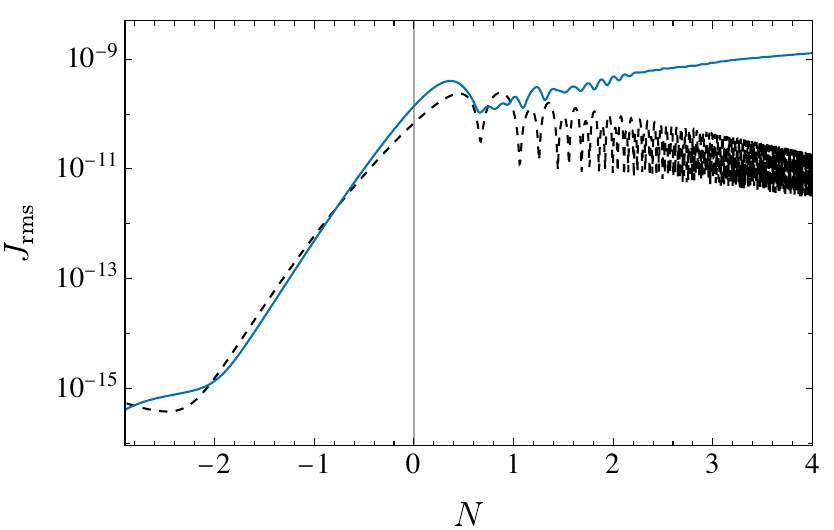}\\
\includegraphics[width=0.45\columnwidth]{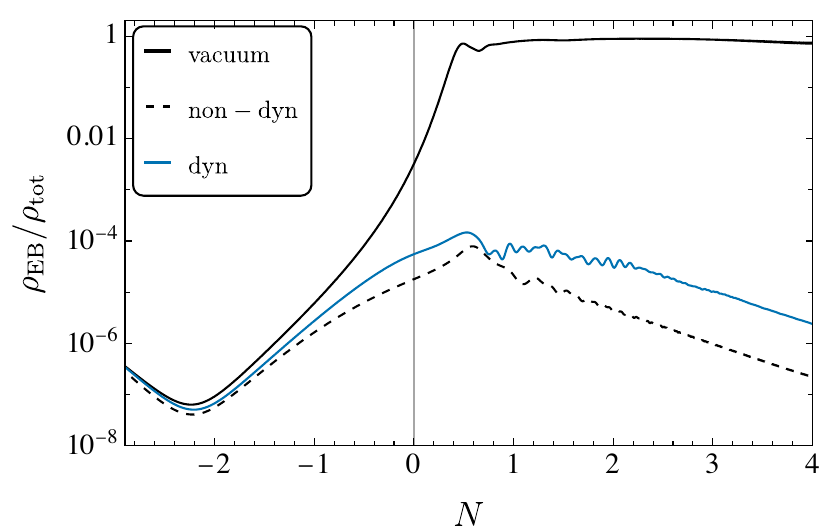}
\includegraphics[width=0.45\columnwidth]{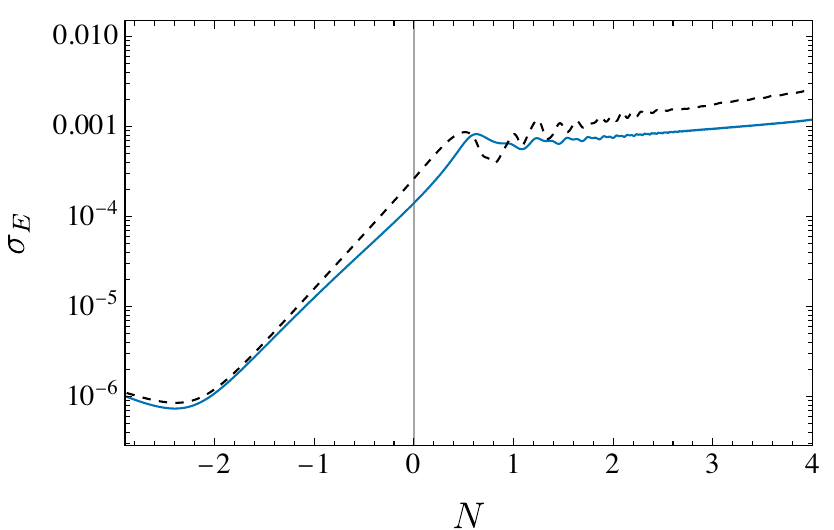}
\end{center}\caption[]{
Comparison of physical ${B}^{\rm ph}_\mathrm{rms}$, ${E}^{\rm ph}_\mathrm{rms}$ and ${J}^{\rm ph}_\mathrm{rms}$ and comoving ${E}_\mathrm{rms}$, ${B}_\mathrm{rms}$, ${J}_\mathrm{rms}$,  $\rho_{EB}/\rho_{\mathrm{tot}}$, $\sigma_E$
for cases with dynamical and non-dynamical current.
The axion-gauge coupling is chosen as $\alpha m_{\rm {Pl}}/f=60 $. 
In the upper-left panel, black curves (solid for $B^{\rm ph}_\mathrm{rms}$, dashed for $E^{\rm ph}_\mathrm{rms}$) show the fields with a non-dynamical current, while blue curves (solid for $B^{\rm ph}_\mathrm{rms}$, dashed for $E^{\rm ph}_\mathrm{rms}$) show the fields with a dynamical current. 
The middle panel is similar to the top panel, but it displays comoving fields ${B}_\mathrm{rms}=a^2{B}^{\rm ph}_\mathrm{rms}$, ${E}_\mathrm{rms}=a^2{E}^{\rm ph}_\mathrm{rms}$ and ${J}_\mathrm{rms}=a^3 {J}^{\rm ph}_\mathrm{rms}$.
In the lower-left panel, the solid black curve shows the EM energy density for the vacuum case (without Schwinger suppression), the dashed blue curve is for the Schwinger case with a non-dynamical current, and the solid blue curve is for the case with a dynamical current. The same color coding applies to the upper-right, middle-right and lower-right panels showing comoving ${J}_\mathrm{rms}$ and $\sigma_E$. The vertical gray line indicates the end of inflation. 
}\label{comp_current_3D}\end{figure}

Since we are evolving the full system on the lattice, there is no added complication in numerically integrating Eq.~\eqref{eq:Jderiv} instead of using Eq.~\eqref{eq:Jintegrated}.
Figure~\ref{comp_current_3D} shows the results of a simulation run, where the axion and EM fields are computed at each point on the lattice, while the current is computed by integrating Eq.~\eqref{eq:Jintegrated}, similarly at each point on the lattice.

From the top panel of Figure~\ref{comp_current_3D} we observe that the evolution of the physical electric and magnetic fields, as well as the current, is similar for both dynamical and non-dynamical current prescriptions. At the end of inflation, however, the dynamical current has an amplitude roughly twice that of the non-dynamical one. Although the dynamical field is somewhat larger, its amplitude is still too small to produce a significant effect. In Section~\ref{supp:Magnetogenesis}, we show that this difference leads to almost indistinguishable present-day magnetic fields. 


The middle panel of Figure~\ref{comp_current_3D} displays the evolution of the comoving electric and magnetic fields and the comoving current. It is worth noting the difference in the late-time evolution of comoving electric and magnetic fields for the dynamical and non-dynamical prescriptions. In Section~\ref{supp:Magnetogenesis} we discuss in detail that, for the late-time evolution, the crucial aspect is conservation of magnetic helicity, where the evolution of the magnetic field is tied to the coherence length (such that the evolution of the magnetic field should always be considered together with that of 
the coherence length). We show that the mean magnetic helicity density is approximately conserved for both treatments of the Schwinger current, dynamical and non-dynamical (see Figure \ref{Fig:B2L}). In particular for the dynamical current, after the end of inflation, the magnetic field increases while the coherence length decreases, with the two effects compensating each other so that the mean magnetic helicity density remains approximately constant.
This allows us to assume that the inverse cascade operates from around the end of inflation for both cases. We emphasize that, within this treatment, the dominant contribution for the present-day values comes from around the peak of both the magnetic field and the coherence length during their generation (see Figure \ref{Fig:Brmsph} and the discussion in Section~\ref{supp:Magnetogenesis}).

The bottom-left panel of Figure~\ref{comp_current_3D} compares the total energy density in the absence of the Schwinger effect (vacuum case, solid black) with the cases including Schwinger current, for both dynamical (blue) and non-dynamical (dashed black) prescriptions. The energy density evolves similarly in both cases, with slightly weaker suppression from Schwinger backreaction in the dynamical case. The bottom-right panel shows the evolution of the conductivities. We find that the electric conductivity, $\sigma_E$, evolves very similarly in both cases, which explains the comparable suppression of the magnetic fields.

\subsection{Magnetogenesis}
\label{supp:Magnetogenesis}

This section presents the computation details of the present-day magnetic field. 
The root mean square value of the physical magnetic field and its comoving coherence length are given by
\begin{equation}
    B^{\rm {ph}}_{\rm{rms}}=\frac{1}{a^2}\sqrt{\int d \log k \, \cdot P_B(k)}, \quad \quad L_c=\frac{\int d \log k \, \cdot k^{-1}\cdot P_B}{\int d \log k \, \cdot P_B}  \, .
\end{equation}


\begin{figure}[]\begin{center}
\includegraphics[width=0.45\columnwidth]{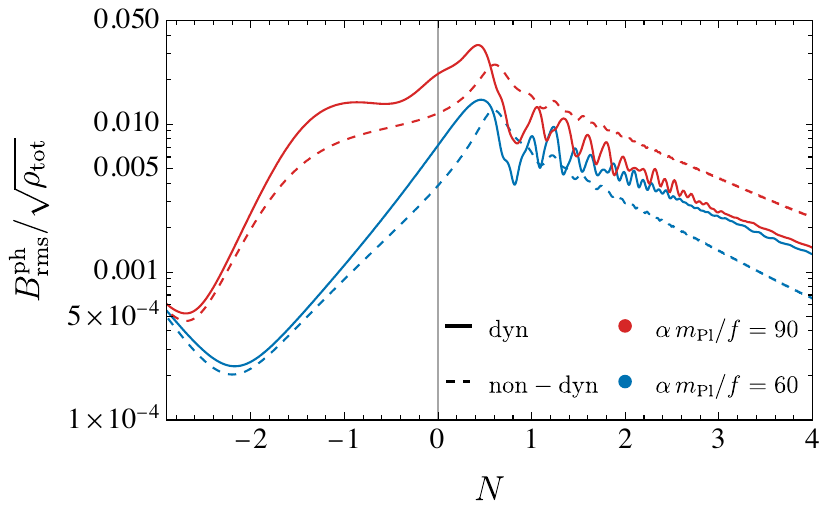}
\includegraphics[width=0.45\columnwidth]{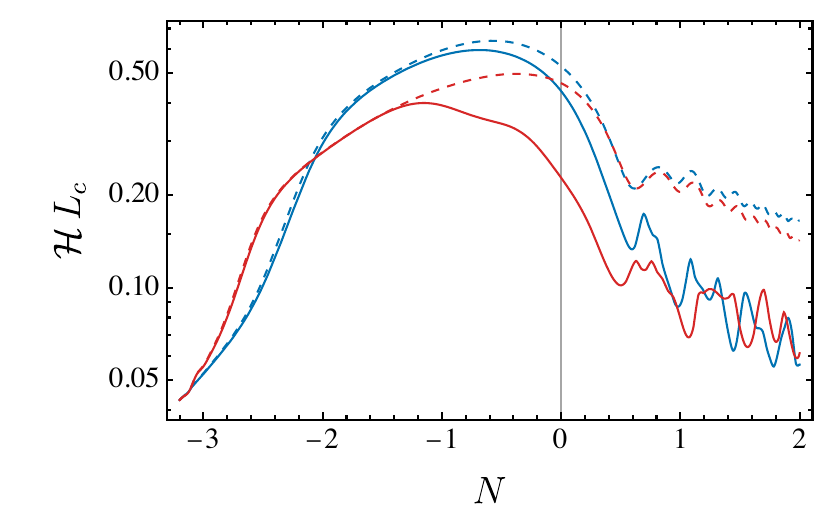}
\end{center}\caption[]{
The magnetic field strength normalized by  the total energy density (left) and the coherence length normalized to the comoving Hubble scale (right) for dynamical (solid) and non-dynamical (dashed) currents. Red curves correspond to $\alpha \, m_{\rm {Pl}}/f=90$ and blue curves to $\alpha \, m_{\rm {Pl}}/f=60$. The vertical gray line indicates the end of inflation. The color-coding is the same for both panels. The dynamical current prescription yields a dimensionless magnetic field strength that is approximately twice as large at the end of inflation, but with a shorter coherence length. Taking into account the nonlinear processing via equations \eqref{eq:BrmsL}, this leads to nearly indistinguishable present-day magnetic fields, which are about two orders of magnitude lower than the lower bound inferred from blazar observations. Figure~\ref{Fig:BLplotmain} illustrates how these results compare with current observational constraints.
}\label{Fig:Brmsph}\end{figure}

After accounting for the nonlinear evolution of the fields, the magnetic field strength and its coherence length at the present epoch are given by (see \cite{Sharma:2024nfu} for details)
\begin{align} \label{eq:BrmsL}
	B_{\rm{rms}}^{\rm{ph}}|_0=9.2 \times 10^{-15}\,{\rm G} \, \sqrt{\frac{\int d \log k \, \cdot P_B}{\rho_{\rm{tot}}}}\left( \frac{10^{-6}m_{\rm {Pl}} }{H}\right)r_{A}^{1/3} \, ,  \quad
	L_c|_0=0.8 \,{\rm pc}\, ({\cal H}L_c)\left( \frac{10^{-6}m_{\rm {Pl}} }{H}\right)r_{A}^{-2/3}  \, ,
\end{align}
where
$r_{A}=\rm{max} (1, {\cal H}L_c/v_A)$ and
 $v_A=\sqrt{B^2/(\rho_{\rm{\chi}}+p_\chi)}$ is the Alfv\'en velocity that characterizes the propagation speed of magnetohydrodynamic (MHD) disturbances in a plasma, where we assume $p_\chi=\rho_{\rm{\chi}}/3$. The definition of the Alfv\'en velocity depends critically on the presence of a conducting medium. Without the Schwinger effect, the Universe close to the end of inflation is essentially vacuum-like, and the standard model plasma only forms after the end of inflation through reheating via the decay of the inflaton field into relativistic particles. 
 In this case, since there is no significant plasma yet, the Alfv\'en velocity is ill-defined. For axion inflation, however, reheating can be effectively instantaneous, which allows for a conducting plasma to form quickly and for the Alfv\'en velocity to become meaningful.

Including the Schwinger effect further changes this picture: the strong gauge fields generated during axion inflation can produce electron-positron pairs via nonperturbative particle creation. This process generates a conducting plasma well before the conventional reheating phase, providing a medium in which MHD waves can propagate. Consequently, the Alfv\'en velocity becomes well-defined earlier, and its value reflects the energy density of the newly created plasma. This distinction is crucial for accurately modeling magnetic field evolution: neglecting the Schwinger effect leads to an incorrect estimate of the plasma contribution, which in turn miscalculates the Alfv\'en velocity and the resulting magnetogenesis dynamics in the early Universe.

Figure \ref{Fig:Brmsph} shows the evolution of the dimensionless magnetic field strength relative to the total energy density and the coherence length normalized to the comoving Hubble scale. The dynamical current produces a dimensionless magnetic field strength approximately twice as large at the end of inflation, but with a smaller coherence length. After accounting for the nonlinear evolution of the fields using equations \ref{eq:BrmsL}, we compute the resulting present-day magnetic field strength and coherence length.  It is clear from Fig.~\ref{Fig:Brmsph} that the dominant contribution comes from around the end of inflation. This is because both the amplitude and the coherence length entering Eqs.~\ref{eq:BrmsL} are dominated by their peak values, which occur around the end of inflation.
Figure \ref{Fig:BLplotmain} compares these results with observational constraints. Due to the universal suppression from the Schwinger backreaction, the results for both coupling constants $\alpha m_{\rm {Pl}}/f = 90$ and $\alpha m_{\rm {Pl}}/f = 60$ are very close to each other, see Table \ref{tab:Bvalues}. Regardless of the Schwinger current description, the resulting magnetic field strength is suppressed by roughly two orders of magnitude, effectively ruling out magnetogenesis in axion inflation.

\begin{table}[t]
\centering
\begin{tabular}{|c|c|c|c|c|c|c|}
\hline
$\alpha m_{\rm {Pl}}/f$ & $B_{\rm{rms}}^{\rm{ph}}|_0$ \text{(vac) [G]} & $B_{\rm{rms}}^{\rm{ph}} |_0$\text{(non-dyn) [G]}   & $B_{\rm{rms}}^{\rm{ph}}|_0$ \text{(dyn) [G]}&  
$L_c|_0$ \text{(vac) [Mpc]} &  $L_c|_0$ \text{(non-dyn) [Mpc]}& $L_c|_0$ \text{(dyn) [Mpc]} \\
\hline
60 & $1.7\times 10^{-15}$ & $7.25\times 10^{-17}$ & $1.35\times 10^{-16}$ & $ 4.1\times10^{-8}$ & $1.65\times 10^{-7}$ & $1.38\times 10^{-7}$\\
\hline
90 & $1.9\times 10^{-14}$ & $2.15\times 10^{-16}$ & $2.92\times 10^{-16}$ & $2.5\times 10^{-7}$ & $1.43\times 10^{-7}$ & $1.28\times 10^{-7}$\\
\hline
\end{tabular}
\caption{Comparison of the present-day magnetic field strength $B_{\rm{rms}}^{\rm{ph}}|_0$ and coherence length $L_c|_0$ for coupling constants $\alpha m_{\rm {Pl}}/f = 90, 60$, considering the case without the Schwinger effect (vacuum case) and with the Schwinger effect using both non-dynamical and dynamical current prescriptions. The values are computed from the power spectrum at the end of inflation. Reference values for the vacuum case are taken from \cite{Sharma:2024nfu}. The observational bound is given by $B_{\rm{bound}}=1.8\times 10^{-17} \text{G}\, \left(L_c/0.2 \text{Mpc}\right)^{-1/2}$ \cite{MAGIC:2022piy}. For a coherence length $L_c=10^{-7} \,\text{Mpc}$, this corresponds to  $B_{\rm{bound}} =2.55\times 10^{-14}$ G. The results including the Schwinger effect are at least two orders of magnitude below this bound, even at strong coupling and regardless of the current description.
}
\label{tab:Bvalues}
\end{table}

It is interesting to consider the effect of the Schwinger backreaction on the helicity of the produced magnetic fields. The magnetic energy and magnetic helicity can be expressed as integrals of the corresponding energy and helicity densities in both position and momentum space \cite{Adshead:2016iae} 
\begin{align}
 E_B(t)&=\frac{1}{V}\int_Vd^3x \frac{\langle B^2 \rangle}{2}=\int_0^{\infty} dk \, {\cal E}_B (k,t),\\
  H_B(t)&=\frac{1}{V}\int_Vd^3x \, \langle A\cdot B \rangle=\int_0^{\infty} dk \, {\cal H}_B (k,t).
\end{align}
It is useful to define the integrated helicity fraction as
 \begin{equation}
     h_B=\frac{H_B}{2L_c E_B},
 \end{equation}
 which is invariant under the transformation between physical and comoving fields.
By construction, the value of the integrated helicity fraction is between zero and one in the absolute values, with the equality holding for maximally helical fields. In the absence of the Schwinger effect and for small axion-gauge couplings, the resulting magnetic power spectrum is effectively maximally helical. This occurs because the gauge-field mode amplified during inflation grows by several orders of magnitude more than the mode that is amplified only during preheating. In the intermediate-to-large coupling regime, however, rescattering effects between the gauge and axion modes become important. In this case, the subdominant helicity is generated through the scattering of the dominant mode off the axion field. This process mainly affects shorter wavelength modes (large wavenumbers), leading to a nearly non-helical field on small scales, while the large-scale modes produced during inflation remain significantly helical \cite{Adshead:2015pva,Adshead:2016iae}. 

\begin{figure}[]\begin{center}
\includegraphics[width=0.5\columnwidth]{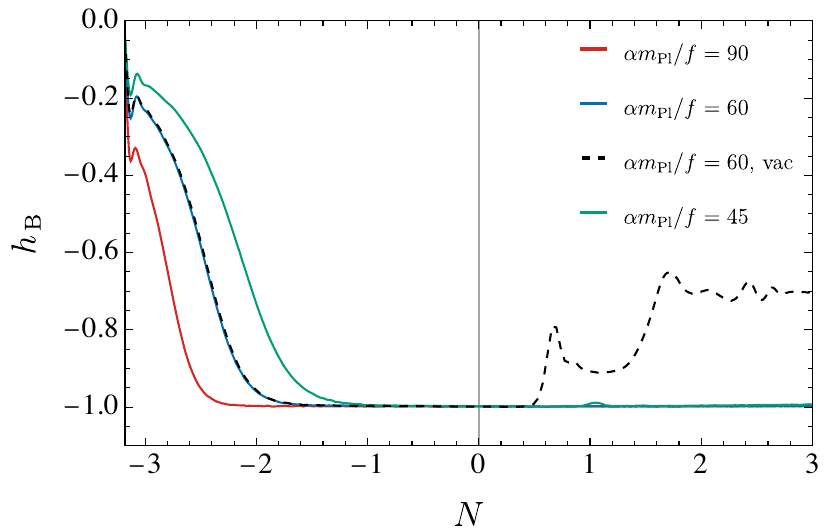}
\end{center}\caption[]{
The integrated
helicity fraction 
as a function of the number of e-folds for $\alpha m_{\rm{Pl}}/f = 90$ (red), $\alpha m_{\rm{Pl}}/f = 60$ (blue) and  $\alpha m_{\rm {Pl}}/f = 45$ (green), including the Schwinger effect. The black dashed curve shows $\alpha m_{\rm {Pl}}/f = 60$ for the vacuum case (without the Schwinger effect). The vertical gray line indicates the end of inflation.  With the Schwinger effect, nonlinear backreaction and rescattering are suppressed, so the subdominant helicity mode is not enhanced, and the magnetic field remains maximally helical throughout reheating. By contrast, in the vacuum case, rescattering generates the opposite helicity, reducing the integrated helicity fraction to about 0.7 in the absolute value.
}\label{Fig:Hel}\end{figure}

The situation changes in the presence of the Schwinger effect. As discussed in the main part of the Letter, the Schwinger effect suppresses nonlinear backreaction, preventing rescattering between the axion and gauge fields. As a result, the magnetic field remains maximally helical throughout reheating, as illustrated in Figure \ref{Fig:Hel}. We show the integrated helicity fraction as a function of time for evolution with the Schwinger effect at couplings $\alpha m_{\rm{Pl}}/f = 90$ (red), $\alpha m_{\rm{Pl}}/f = 60$ (blue) and  $\alpha m_{\rm{Pl}}/f = 45$ (green). In all cases, the magnetic field becomes fully helical shortly after the start of the simulation and retains its maximal helicity until the end of the simulation. For comparison, we also show the evolution with $\alpha m_{\rm{Pl}}/f = 60$ (black dashed) in the vacuum case (without the Schwinger effect). Here, the magnetic field initially becomes helical, but strong nonlinear backreaction and rescattering subsequently generate the growth of opposite polarization, reducing the integrated helicity fraction to an absolute value of approximately 0.7. This difference is crucial for the subsequent evolution of the magnetic fields \cite{Brandenburg:2025ccv}.

\begin{figure}[]\begin{center}
\includegraphics[width=0.5\columnwidth]{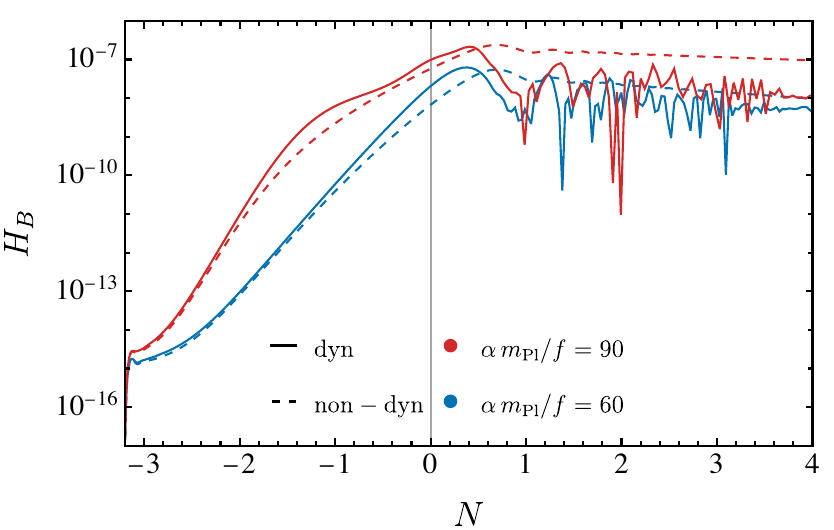}
\end{center}\caption[]{The evolution of the magnetic helicity density (absolute value) expressed in comoving fields for $\alpha m_{\rm{Pl}}/f = 90$ (red), $\alpha m_{\rm{Pl}}/f = 60$ (blue), and for both dynamical (solid) and non-dynamical (dashed) treatments of the Schwinger current. We see that the helicity density is approximately constant after the end of inflation for both dynamical and non-dynamical cases. 
}\label{Fig:B2L}\end{figure}

Another crucial aspect is the conservation of magnetic helicity. The helicity density is approximately conserved \cite{ Berger:1984izr, Berger:1984}
\begin{equation}
    H_{B}=h_B\, B_{\rm {rms}}^2L_c \approx \rm{const}.
\end{equation}
As a result, helicity conservation imposes a non-trivial constraint on the subsequent evolution of the magnetic field.
In particular, as the system evolves and magnetic energy is dissipated, the coherence length must increase in order to maintain approximately constant helicity. This leads to an inverse transfer of power from small to large scales, with $B_{\rm {rms}}^2$ decreasing while $L_c$ grows such that their product remains nearly constant. Consequently, even if magnetic fields are initially generated on small scales, helicity conservation enables the efficient build-up of large-scale coherent fields at later times. We show the evolution of the helicity density in Figure \ref{Fig:B2L} for two coupling constants $\alpha m_{\rm{Pl}}/f = 90$ (red), $\alpha m_{\rm{Pl}}/f = 60$ (blue), and for both dynamical (solid) and non-dynamical (dashed) treatments of the Schwinger current. We see that initially the helicity fraction grows as expected, but after the end of inflation it is approximately conserved for both treatments of the Schwinger current.

We do not expect exact conservation of helicity, since the system is not in the ideal MHD limit. In particular, finite conductivity, the presence of Schwinger pair production and the inflaton introduce non-ideal effects that can lead to helicity dissipation or increase. Nevertheless, the approximate conservation observed here indicates that these effects are not efficient enough to significantly alter the helicity budget at late times. 
This suggests that the subsequent evolution of the magnetic field may proceed via an inverse transfer, with a gradual increase of the coherence length while maintaining nearly constant $B_{\rm {rms}}^2 L_c$, as assumed in equations for magnetic field evolution \eqref{eq:BrmsL}.

We must stress here, that even though we are able to follow the simulation of the inflaton-photon-Scwhinger plasma system for several $e$-folds after the end of inflation, simulation does not include the MHD effects. In the next Section~\ref{supp:Inhomogeneous} we outline possible implications of them.

\subsection{Inhomogeneous energy density and bulk motions}
\label{supp:Inhomogeneous}

\begin{figure}[]\begin{center}
\includegraphics[width=0.5\columnwidth]{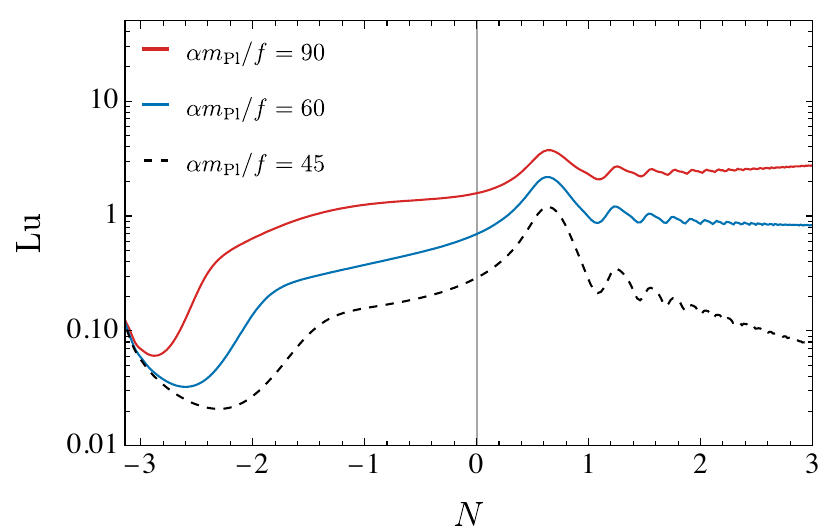}
\end{center}\caption[]{
The Lundquist number $\mathrm{Lu}=v_\mathrm{A} L_c\sigma_E$ versus the number of $e$-folds $N$ for $\alpha m_{\rm {Pl}}/f=90$ (red),
$60$ (blue), and $45$ (black dashed). The vertical gray line indicates the end of inflation. The Lundquist number exceeds unity only very close to or after the end of inflation, suggesting that any onset of plasma motion may become relevant at this late stage.}
\label{pcomp_GW3N_Rey}\end{figure}

For completeness, we also note that the Schwinger pair plasma will not remain at rest, but it can be accelerated by the Lorentz force, $\bm{J}\times\bm{B}$.
This leads to some of the EM energy being converted into kinetic energy at the rate $\langle\bm{u}\cdot(\bm{J}\times\bm{B})\rangle$
where $\bm{u}$ is the bulk velocity of the plasma \cite{BP23}. A useful measure of the importance of nonlinear effects is the magnetic Reynolds number.
However, in the absence of a well-defined bulk flow velocity, this quantity cannot be meaningfully constructed. Instead, the natural velocity scale in the system is the Alfv\'en speed  $v_\mathrm{A}$, and the relevant dimensionless parameter becomes the Lundquist number
\begin{equation}
    \mathrm{Lu}=v_\mathrm{A} L_c\sigma_E, \label{eq:Lu}
\end{equation} which plays an analogous role in characterizing the relative importance of advection and diffusion of the magnetic field.
When the Lundquist number becomes large, we can no longer ignore the production of inhomogeneities and the resulting inhomogeneous bulk motions of the plasma.
The evolution of $\mathrm{Lu}$ is shown in Fig.~\ref{pcomp_GW3N_Rey} for three values of the coupling strength $\alpha m_{\rm {Pl}}/f$.
Evidently, for large axion-gauge couplings, nonlinearity and hydromagnetic effects can become important.
However, we observe that the Lundquist number exceeds unity only near or after the end of inflation.
This would mean that part of the generated EM energy is being converted into kinetic energy,
which might diminish the magnetic field even further. Consequently, primordial magnetogenesis from axion inflation can become even more challenging, leaving the main conclusion of our paper intact.

The physical effect of bulk motions is that they affect the EM field and re-organize it to form larger scale
structures while preserving the magnetic helicity.
The resulting energy spectra then exhibit a phenomenon known as an inverse cascade \citep{BEO96}.
To account for the resulting inhomogeneities and the bulk motions of plasma,
Eq.~\eqref{eq:fermion_energy} must be amended to include extra divergence terms that would vanish after volume averaging.
Phenomenologically, the governing equations for the \textit{comoving} fermion energy density, $\tilde{\rho}_{\chi}=a^4 \rho_{\rm{ph},\chi}$, take the form
\begin{equation}\label{eq:inhom_fermion_energy}
\partial_\tau\tilde{\rho}_{\chi}+\bm{\nabla}\cdot(\textstyle{\frac{4}{3}}\tilde{\rho}_\chi\bm{u})=\bm{J}\cdot \bm{E}
+2\tilde{\rho}_\chi\nu\bm{\mathsf{S}}^2+\bm{\nabla}\cdot(\tilde{\rho}_\chi\kappa\bm{\nabla}\tilde{\rho}_\chi),
\end{equation}
where $\bm{u}$ is the velocity of the bulk motions of fermions, which obeys the usual momentum equation 
\begin{equation}\label{eq:bulk_motion}
    \textstyle{\frac{4}{3}}\tilde{\rho}_\chi\left(\partial_\tau\bm{u}+\bm{u}\cdot\bm{\nabla}\bm{u}\right)=-\textstyle{\frac{1}{3}}\bm{\nabla}\tilde{\rho}_\chi+\bm{J}\times\bm{B}
    +\bm{\nabla}\cdot(2\tilde{\rho}_\chi\nu\bm{\mathsf{S}}),
\end{equation}
where $\nu$ is the viscosity and $\bm{\mathsf{S}}$ is the rate-of-strain tensor with the components
$\mathsf{S}_{ij}=(\partial_i u_j+\partial_j u_i)/2-\delta_{ij}\partial_k u_k/3$.
The  term $2\tilde{\rho}_\chi\nu\bm{\mathsf{S}}^2$ on the right-hand side of Eq.~\eqref{eq:inhom_fermion_energy}
denotes viscous heating and $\kappa$ is a radiative diffusion coefficient, which we expect to be comparable to the viscosity.
Since we work here with comoving fermion energy density, the $4 {\cal H} \rho_{\chi}$ term from Eq.~\eqref{eq:fermion_energy} has disappeared.

The bulk motions also affect the current density, which then obeys the equation
\begin{equation}
    \bm{J} = \sigma_E (\bm{E}+\bm{u}\times\bm{B}) + \sigma_B \bm{B}\, . \label{eq:mixed_bulk}
\end{equation}
The $\bm{u}\times\bm{B}$ term is the electromotive force and leads to a feedback of the fermion velocity onto the EM field.
The nonlinear interaction between $\bm{u}$ and $\bm{B}$ leads to a turbulent cascade of energy that is finally dissipated through the first two terms
on the right-hand side of Eq.~\eqref{eq:inhom_fermion_energy}; see Ref.~\cite{Kahniashvili+17} for earlier work on these aspects.
The detailed investigation of the resulting turbulence is, however, beyond the scope of the present paper and presents a fruitful avenue for future work.

\end{document}